\documentclass[amsmath,aps,pra]{revtex4}
\usepackage{graphicx}
\usepackage{times}
\usepackage{bm}

\begin{document}
\title{Two-photon atomic level widths at finite temperatures}
\author{T. Zalialiutdinov$^{1}$, A. Anikin$^{1}$, D. Solovyev$^1$ }

\affiliation{ 
$^1$ Department of Physics, St. Petersburg
State University, Petrodvorets, Oulianovskaya 1, 198504,
St. Petersburg, Russia
\\
$^2$ Petersburg Nuclear Physics Institute, 188300, Gatchina, St. Petersburg, Russia}

\begin{abstract}
The thermal two-photon level broadening of the excited energy levels in the hydrogen and H-like helium is evaluated via the imaginary part of thermal two-loop self-energy correction for bound electron. All the derivations are presented in the framework of rigorous quantum electrodynamic theory at finite temperatures and are applicable for the H-like ions. On this basis, we found a contribution to the level broadening induced by the blackbody radiation which is fundamentally different from the usual line broadening caused by the stimulated two-photon decay and the Raman scattering of thermal photons. Numerical calculations of the two-loop thermal correction to the two-photon width for the $2s$ state in hydrogen and singly ionized helium atoms show that the effect could significantly exceed the higher-order relativistic and radiative QED corrections commonly included in the calculations. In addition, the thermal two-loop self-energy correction significantly exceeds the "ordinary" stimulated one-photon depopulation rate at the relevant laboratory temperatures. In this work, detailed analysis and the corresponding comparison of the effect with the existing laboratory measurements in H-like ions are carried out. 
\end{abstract}

\maketitle
\section{Introduction}
\label{intro}

During the last decades two-photon processes became of high interest in fundamental investigations of field theories, astrophysics, laboratory experiments, constructing of atomic clocks, chemistry and biology~\cite{example1,example2,example3,example4,example5,example6,example7,chluba1}. Since the early days of quantum mechanics, a special role was assigned to the two-photon decay of $2s$ state in hydrogen atom~\cite{goppert}. Recent accurate measurements of the temperature and polarization distribution of the cosmic microwave background (CMB) renewed interest in this process~\cite{planck,chluba1}. 
The modern theory of the cosmological recombination starts from works by Zel'dovich, Kurt, and Sunyaev~\cite{zeld} and Peebles~\cite{peebles}. In particular, the $2s \rightarrow 1s + 2\gamma  (\mathrm{E1})$ decay rate in hydrogen was found to be the main channel within the bound-bound transitions for the radiation escape from the matter and formation of CMB. Hence the present properties of the CMB are strongly depend on the particular qualities of the two-photon processes during the cosmological recombination epoch. In addition to the transition $2s \rightarrow 1s + 2\gamma (\mathrm{E1})$, no less attention is paid to the two-photon decays of excited states with principal quantum number $n> 2$, whose total contribution to the ionization fraction of primordial plasma reaches the percent level and exceeds the accuracy of CMB measurements~\cite{chluba1,hirata1, solovyevhighn}. 

However, the study of two-photon transitions for excited states is complicated by the crucial difference between the decays of $nl$ $(n > 2)$ and $2s$ atomic levels. This difference is determined by the presence of cascade transitions as the dominant decay channels of the excited levels which are absent in the case of $2s$ level. In connection with the presence of a cascade channel (resonant transitions), the problem of the separating of nonresonant two-photon emission (resulting in the immediate radiation escape from the matter) arose in astrophysical studies. This question was studied within the quantum mechanical approach in a number of works, see for example \cite{chluba1,hirata1}.
 Within the framework of QED theory the ambiguity of such separation was demonstrated in \cite{ambiguity1,ambiguity2,ambiguity3}, while an alternative approach to obtaining a "pure" two-photon contribution based on an evaluation of the imaginary part of the two-loop self-energy of bound electron was proposed in \cite{jent1,jent2,jent3,surzhykov}. According to this "alternative" approach the found contributions were called "two-photon widths" since in the absence of cascade emission (case of the $2s$ state in a hydrogen atom) it coincides with the two-photon transition rate, \cite{jent1,jent2,jent3,surzhykov}. However, within the framework of the Line Profile Approach \cite{olegreports} it was shown that the imaginary part of the two-loop radiative level shift does not coincide with the two-photon decay rate for higher states and should be considered only as a radiative correction to the level widths~\cite{ourtwoloop}, see also \cite{physrep2018}. The study of radiative correction of this type (but in the thermal case) according to the "alternative" method suggested in \cite{jent1,jent2,jent3,surzhykov} is the main purpose of present work and we will use the designation "two-loop width" assuming the imaginary part of two-loop self-energy radiative correction to the energy level.

Besides the spontaneous decays, the corresponding transitions induced by the blackbody radiation (BBR) are also of particular interest. Accounting for the induced level broadening leads to an additional correction to CMB properties. A comprehensive analysis of the induced two-photon transitions in recombination processes based on quantum mechanical approach has been the subject of discussion in~\cite{chlubaind,hirataind,kholupenkoind}. However, the necessity to use the quantum electrodynamics (QED) theory was recently demonstrated in~\cite{solovyev2015,zalialiutdinov2017,zalialiutdinov2018,zalialiutdinov2019}, see also \cite{solovyev2019,onephoton}. Since the CMB has an almost Planck spectrum, the atomic line broadening can be described within the framework of quantum electrodynamics (QED) for bound states at finite temperatures~\cite{solovyevarxiv2019}.
 
The particular attention can be paid to the analysis of CMB properties and the corresponding determination of the $2s$ state lifetime in a hydrogen atom with an 8\% error which is much better than in any existing laboratory experiments~\cite{2slife}. Nonetheless, despite a special astrophysical role of the $2s$ state in hydrogen atom, its importance is even more significant for the laboratory spectroscopic experiments. Being especially metastable, this state allows precision measurements of various transition frequencies with an accuracy reaching $10^{-13}$ of relative magnitude in hydrogen, pursuing to improve optical standards of frequency, accurate determination of physical constants and testing fundamental interactions in hydrogen and H-like atomic systems. To accomplish this intention, an accurate theoretical calculation of the two-photon widths of metastable states is also of experimental importance. The results of laboratory measurements of the $ 2s $ state lifetime in a hydrogen atom can be found in~\cite{cesar,connel,kruger}, and for the singly ionized helium in~\cite{lipeles}. The experimental data for the H-like highly charged ions (HCI) can be found, for example, in~\cite{prior, cocke, marrus, hinds, gould, cheng,dunford}. 
Generally speaking, experiments on measuring the natural widths of energy levels are incredibly complicated. Improving the accuracy of an experiment requires controlling the impact of physical conditions, such as the influence of external fields, Doppler broadening, evaluation of QED corrections, and such tiny effects as the Stark shifts and level broadening induced by BBR~\cite{farley,gallagher}. 

The formalism of thermal QED theory for bound states developed in~\cite{solovyevarxiv2019} allows one to take into account more complex effects revealing the impact of thermal environment on atomic systems~\cite{onephoton,solovyev2019}. Following this theory, we investigate the two-photon level broadening caused by the "heat bath" employing the "alternative approach" \cite{jent1,jent2,jent3,surzhykov} to evaluate the imaginary part of the two-loop self-energy corrections for a bound electron. The "heat bath" acting on the atomic system implies an environment described by blackbody radiation, i.e. the photon field distributed according to Planck's law. 

The paper is organized as follows. In section~\ref{sectionA} the general equations for the induced two-photon transitions and Raman scattering of thermal photons are given. The derivation of two-photon decay widths at finite temperatures within the two-loop approach is given in section~\ref{sectionB} and the corresponding expressions are presented in section~\ref{sectionC} within the nonrelativistic limit. All the derivations are presented in the framework of rigorous quantum electrodynamics theory at finite temperatures and are applicable for the H-like ions. The results of numerical calculations of the thermal two-photon decay widths for hydrogen and singly ionized helium atoms and their comparison with the existing laboratory measurements are discussed in section~\ref{theend}. Throughout the paper we use the relativistic units $ \hbar=m_{e}=c=1 $ ($ m_{e} $ is the electron rest mass, $c$ is the speed of light and $\hbar$ is the reduced Planck constant).

\section{Induced two-photon transitions and Raman scattering in hydrogen}
\label{sectionA}

In this section, a description of the two-photon transition rates in hydrogen in the presence of the blackbody radiation (BBR) is given briefly. In the absence of external fields, only spontaneous decays of atomic states are possible. Isotropic external radiation, such as BBR, leads to additional level broadening due to the processes of thermal radiation, absorption, and Raman scattering. In the nonrelativistic limit and electric dipole approximation the total rate of two-photon decay $ a\rightarrow b+2\gamma(\mathrm{E1}) $ ($ a(b) $ denotes the standard set of quantum numbers $ n_{a(b)}l_{a(b)} $, where $ n_{a(b)} $ is the principal quantum number of the state $ a(b) $ and $ l_{a(b)} $ is the corresponding orbital angular momentum) in hydrogen-like atoms after the integration over photon directions, summation over photon polarizations, averaging over the projections $m_{a}$ of initial state and summation over projection of final state $m_{b}$, see~\cite{physrep2018}, transforms to 
\begin{eqnarray}
\label{tot}
W^{2\gamma,\mathrm{tot}}_{ab}=\frac{1}{2}\int\limits_{0}^{\omega_{0}}dW^{2\gamma,\mathrm{tot}}_{ab}.
\end{eqnarray}
Here $ dW^{2\gamma,\mathrm{tot}}_{ab}=dW^{2\gamma,\mathrm{spon}}_{ab}+dW^{2\gamma,\mathrm{ind}}_{ab} $, $ \omega_{0}=|E_{a}-E_{b}| $ is the transition energy, and the differential spontaneous decay rate is expressed by
\begin{eqnarray}
\label{twophoton2}
dW^{2\gamma,\mathrm{spon}}_{ab}=\frac{8e^4}{9\pi}
\frac{\omega^3 (\omega_{0}-\omega)^3}{2l_{a}+1}\sum\limits_{m_am_b}
\left|\sum\limits_{n}
\left(
\frac{\langle b|\textbf{r}|n\rangle \langle n|\textbf{r}|a\rangle}{E_{n}-E_{a}+\omega} 
+
\frac{\langle b|\textbf{r}|n\rangle \langle n|\textbf{r}|a\rangle}{E_{n}-E_{b}-\omega}
\right)
\right|^2
d\omega
.
\end{eqnarray}

The differential induced decay rate corresponding to the emission process can be obtained in the form:
\begin{eqnarray}
\label{ind}
dW^{2\gamma,\mathrm{ind}}_{ab} =
\frac{8e^4}{9\pi}
\frac{\omega^3 (\omega_{0}-\omega)^3}{2l_{a}+1}\sum\limits_{m_am_b}
\left|\sum\limits_{n}
\left(
\frac{\langle b|\textbf{r}|n\rangle \langle n|\textbf{r}|a\rangle}{E_{n}-E_{a}+\omega} 
+
\frac{\langle b|\textbf{r}|n\rangle \langle n|\textbf{r}|a\rangle}{E_{n}-E_{b}-\omega}
\right)
\right|^2
\\
\nonumber
\times
(n_{\beta}(\omega)+n_{\beta}(\omega_{0}-\omega)+n_{\beta}(\omega)n_{\beta}(\omega_{0}-\omega))
d\omega
.
\end{eqnarray}
In Eq.~(\ref{ind}) $ n_{\beta}(\omega)=(e^{\beta\omega}-1)^{-1} $ represents the Planck distribution function giving the mean occupation number of photons in the BBR field, $ \beta=(k_B T)^{-1} $, $ T $ is the radiation temperature and $ k_B  $ is the Boltzmann constant. 
In the BBR field, the two-photon absorption process $ a +2\gamma(\mathrm{E1})\rightarrow b $  should also be taken into account. The corresponding differential rate reduces to the expression:
\begin{eqnarray}
\label{abs}
dW^{2\gamma,\mathrm{abs}}_{ab} =
\frac{8e^4}{9\pi}
\frac{\omega^3 (\omega_{0}-\omega)^3}{2l_{a}+1}\sum\limits_{m_am_b}
\left|\sum\limits_{n}
\left(
\frac{\langle b|\textbf{r}|n\rangle \langle n|\textbf{r}|a\rangle}{E_{n}-E_{a}-\omega} 
+
\frac{\langle b|\textbf{r}|n\rangle \langle n|\textbf{r}|a\rangle}{E_{n}-E_{b}+\omega}
\right)
\right|^2
n_{\beta}(\omega)n_{\beta}(\omega_{0}-\omega)
d\omega
.
\end{eqnarray}

In the case when $n_{a} = n_{b}\pm 1$, there are no cascade transitions (the energy denominators in these expressions are not equal to zero), the frequency distributions given by Eqs.~(\ref{twophoton2})-(\ref{abs}) are regular and the integral Eq.~(\ref{tot}) is convergent. The results of $dW^{2\gamma,\mathrm{ind}}_{ab}$ evaluation, Eq.~(\ref{ind}), for $ 2s\rightarrow 1s+2\gamma(\mathrm{E1}) $ transition at different temperatures in hydrogen atom (H) and singly ionized helium (He$^{+}$) are given in Table~\ref{tab1}, where the total contribution (spontaneous plus induced) is given also.
\begin{table}
\caption{The BBR induced $ W^{2\gamma,\mathrm{ind}}_{2s1s}$ and total two-photon transition rates $ W^{2\gamma,\mathrm{tot}}_{2s1s}$ (in s$ ^{-1} $) at different temperatures $ T $ (in Kelvin) in H and He$^{+}$ atoms. The nonrelativistic values of spontaneous $ 2s\rightarrow 1s +2\gamma(\mathrm{E1}) $ transition rate in vacuum are $W^{2\gamma,\mathrm{spon}}_{2s1s}=8.229352$ s$ ^{-1} $ and $5.266785\times 10^2$ s$ ^{-1} $~\cite{goldmandrake} for H and He$^{+}$, respectively.}
\begin{tabular}{l c c c c c c c c}
\hline
\hline
atom & $T$ & 77 & 300 & 1000 & 3000  & 5000  & 10$ ^{4} $  \\
\hline
H &$ W^{2\gamma,\mathrm{ind}}_{2s1s} $ & $1.358\times 10^{-4}$ & $2.028\times 10^{-3}$ & $2.151\times 10^{-2}$ &  $1.731\times 10^{-1}$ & $4.389\times 10^{-1}$   & $1.467$ \\
  &$ W^{2\gamma,\mathrm{tot}}_{2s1s}$  & $8.229 $    & $8.231 $     &$ 8.251$     &$ 8.402$      &$ 8.668$     & $9.697$ \\
\hline
\hline
$\mathrm{He}^{+}$ & $ W^{2\gamma,\mathrm{ind}}_{2s1s}$ & $5.438\times 10^{-4}$ & $8.243\times 10^{-3}$ &$ 9.044\times 10^{-2}$ & $ 7.868\times 10^{-1}$ & $2.118\times 10^{-1}   $ & $7.893$ \\
& $ W^{2\gamma,\mathrm{tot}}_{2s1s}$ & $5.267\times 10^{2}$  & $5.267\times 10^{2}$  & $5.268\times 10^{2}$  & $5.275\times 10^{2} $  &$ 5.288\times 10^{2}$     & $5.346\times 10^{2}$\\
\hline
\hline
\end{tabular}
\label{tab1}
\end{table}

\begin{figure}[hbtp]
\caption{Differential transition rate $ dW_{2s1s}^{\mathrm{tot}}(\omega) $ in s$ ^{-1} $ for the $ 2s\rightarrow 1s +2\gamma(\mathrm{E1}) $ transition in H atom in the presence of BBR field with temperature $ T $ (in Kelvin). Bold line corresponds to the transition rate in the absence of BBR field (vacuum).}
\centering
\includegraphics[scale=0.8]{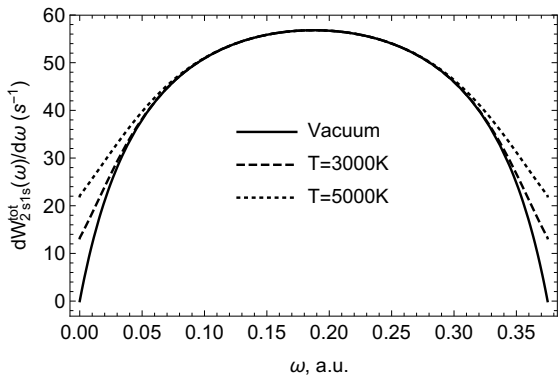}
\label{fig1}
\end{figure}

The frequency distributions for this transition in H are shown in Fig.~\ref{fig1}, where the contribution of the induced transition depending on temperature is observed visually. Calculations shows that the contribution, arising from the cross product $ n_{\beta}(\omega)n_{\beta}(\omega_{0}-\omega) $ in Eq.~(\ref{ind}) and interpreting as interference of two thermal photons, is negligible up to the temperatures of the order of $ T=10^4 $ K. 

In addition to the usual induced emission/absorption transitions, the BBR-stimulated Stokes and anti-Stokes Raman scattering process $ a+\gamma(\mathrm {E1})\rightarrow b +\gamma'(\mathrm{E1}) $ should be also considered. Its differential rate can be obtained within the same methodology~\cite{berest}. The result is
\begin{eqnarray}
\label{raman}
dW^{2\gamma,\mathrm{ram}}_{ab} =
\frac{8e^4}{9\pi}
\frac{\omega^3 (\omega_{0}+\omega)^3}{2l_{a}+1}\sum\limits_{m_am_b}
\left|\sum\limits_{n}
\left(
\frac{\langle b|\textbf{r}|n\rangle \langle n|\textbf{r}|a\rangle}{E_{n}-E_{a}-\omega} 
+
\frac{\langle b|\textbf{r}|n\rangle \langle n|\textbf{r}|a\rangle}{E_{n}-E_{b}+\omega}
\right)
\right|^2
\\
\nonumber
\times
(n_{\beta}(\omega)+n_{\beta}(\omega)n_{\beta}(\omega_{0}+\omega))
d\omega
,
\end{eqnarray}
where for Stokes process $ \omega_{0}=E_{a}-E_{b}<0 $ and for anti-Stokes $ \omega_{0}=E_{a}-E_{b}>0 $. The corresponding total rate is defined by  $W^{2\gamma,\mathrm{AS-ram}}_{ab}=\int\limits_{0}^{\infty}dW^{2\gamma,\mathrm{ram}}_{ab}$ and $W^{2\gamma,\mathrm{S-ram}}_{ab}=\int\limits_{|\omega_{0}|}^{\infty}dW^{2\gamma,\mathrm{ram}}_{ab}$ for anti-Stokes and Stokes processes, respectively.

\begin{figure}[hbtp]
\caption{Differential transition rate $ dW_{2s1s}^{\mathrm{ram}}(\omega) $ in s$ ^{-1} $ for the Raman scattering of BBR photons on H atom in the process $ 2s + \gamma(\mathrm{E1})\rightarrow 1s + \gamma(\mathrm{E1}) $. The BBR field temperature is denoted as $ T $ (in Kelvin). Sharp peaks correspond to resonances in Eq.~(\ref{raman}).}
\centering
\includegraphics[scale=0.8]{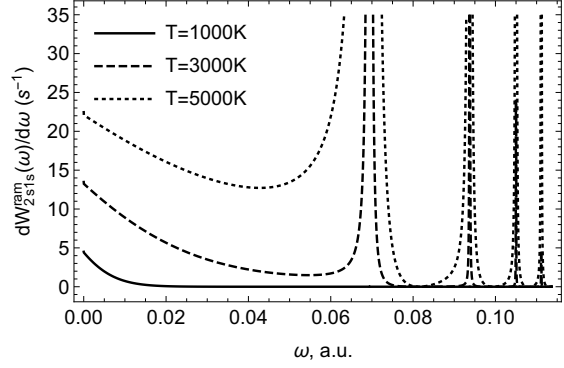}
\label{fig2}
\end{figure}

For the hydrogen atom with the fixed $ a=2s $ and $ b=1s $ states the corresponding distribution of anti-Stokes scattering rate is shown in Fig.~\ref{fig2}.
In particular, from Fig.~\ref{fig2} it follows that the induced cascade contributions in the Raman scattering become significant with increasing of temperature.

\section{Two-loop electron self-energy with one and two thermal loops}
\label{sectionB}

Until now, the well-known effects arising in the BBR field were considered. The results expressed by Eqs.~\ref{twophoton2}-\ref{raman} can be easily obtained within the framework of the quantum mechanical approach. The same can be found within the rigorous QED theory in approximation of zero level widths of bound states. Nevertheless, the application of QED theory can be used to identify new effects, see for example \cite{solovyev2015}. In particular, within the framework of the QED perturbation theory for a bound electron, such self-energy radiative corrections as the one-loop (second-order in the coupling constant), two-loop (fourth-order in the coupling constant) and etc can be treated as a sequential contributions. In the absence of external fields and cascades, the imaginary parts of these corrections give the one-photon width $ \Gamma_{a}^{1\gamma} $, the two-photon width $ \Gamma_{a}^{2\gamma} $, and etc, respectively. Then the total level width $ \Gamma_{a} $ of atomic state $ a $ can be presented by the infinite series
\begin{eqnarray}
\label{zerot}
\Gamma_{a}=\Gamma_{a}^{1\gamma}+\Gamma_{a}^{2\gamma}+\dots
.
\end{eqnarray}

In turn, the natural one-photon width $ \Gamma_{a}^{1\gamma} $ is equal to the sum of the one-photon transition rates to lower levels. In the nonrelativistic limit it can be written as follows
\begin{eqnarray}
\Gamma_{a}^{1\gamma}=\frac{4e^2}{3}\frac{1}{2l_{a}+1}\sum\limits_{b<a}\sum\limits_{m_am_b}\omega_{ab}^3|\langle b | \textbf{r}| a \rangle |^2,
\end{eqnarray}
where $ \omega_{ab}=E_{a}-E_{b} $. Recently, the two-photon width $ \Gamma_{a}^{2\gamma} $ was evaluated within the relativistic adiabatic QED theory~\cite{ourtwoloop}, where the expression for $ \Gamma_{a}^{2\gamma} $ was also obtained in nonrelativistic limit:
\begin{eqnarray}
\label{gammatwophoton}
\Gamma_{a}^{2\gamma} =
\frac{4e^4}{9\pi}\frac{1}{2l_{a}+1}\lim_{\eta\rightarrow 0}\mathrm{Re}\sum\limits_{b<a}\sum\limits_{m_am_b}\int\limits_{0}^{\omega_{ab}}
\omega^3 (\omega_{ab}-\omega)^3
\sum\limits_{nn'}
\left(
\frac{\langle b|\textbf{r}|n\rangle \langle n|\textbf{r}|a\rangle}{E_{n}-E_{a}+\omega +\mathrm{i}\eta} 
+
\frac{\langle b|\textbf{r}|n\rangle \langle n|\textbf{r}|a\rangle}{E_{n}-E_{b}-\omega+\mathrm{i}\eta}
\right)
\\\nonumber
\times
\left(
\frac{\langle b|\textbf{r}|n'\rangle^* \langle n'|\textbf{r}|a\rangle^*}{E_{n'}-E_{a}+\omega+\mathrm{i}\eta} 
+
\frac{\langle b|\textbf{r}|n'\rangle^* \langle n'|\textbf{r}|a\rangle^*}{E_{n'}-E_{b}-\omega+\mathrm{i}\eta}
\right)
.
\end{eqnarray}
This result was obtained at first in~\cite{jent1}. It is worth noting that the expression (\ref{gammatwophoton}) coincides with Eq.~(\ref{twophoton2}) for the two-photon decay rate only in the absence of resonant energy denominators. Then the imaginary infinitesimal part $ \mathrm{i}\eta $ ($\eta$ is the adiabatic parameter) in Eq.~(\ref{gammatwophoton}) can be omitted and the product of two terms in parentheses is equal to the square modulus making $ \Gamma_{a}^{2\gamma} $ the same with spontaneous decay rate $ W_{a}^{2\gamma,\mathrm{spon}} $. 

The situation is different for the two-photon transitions with cascades, i.e. when the presence of resonant intermediate states in the sum over $ n $ in Eq.~(\ref{gammatwophoton}) leads to the divergent contributions. In this case, the transition rate Eq.~(\ref{twophoton2}) should be regularized in the vicinity of resonances. The regularization procedure of multiphoton transition amplitudes within the framework of QED theory can be found, for example, in~\cite{qedandqm}. As a result the corresponding level widths arise in the divergent contributions that leads to the regular expression.

In contrast, there is no need to regularize Eq.~(\ref{gammatwophoton}), see~\cite{jent1,jent2,jent3}. In particular, in~\cite{jent1} it was demonstrated that the integral
\begin{eqnarray}
\label{intjent}
\lim\limits_{\eta\rightarrow 0}\mathrm{Re}\int\limits_0^1 d\omega\left(\frac{1}{a-\omega +\mathrm{i}\eta}\right)^2=\frac{1}{a(a-1)}+O(\eta^2)
\end{eqnarray} 
is finite, when the limit is taken after integration over frequency $ \omega $ (here assumed that $ 0<a<1 $). With that the integral, arising in the expression for the two-photon transition rate or Raman scattering rate, is divergent when $ \eta\rightarrow 0 $:
\begin{eqnarray}
\int\limits_0^1 d\omega\left|\frac{1}{a-\omega +\mathrm{i}\eta}\right|^2=\frac{\pi}{\eta}+\frac{1}{a(a-1)}+O(\eta^2)
\end{eqnarray} 
The integration method corresponding to Eq.~(\ref{intjent}) was also justified in~\cite{ourtwoloop} within the adiabatic S-matrix formalism. However, in opposite to \cite{jent1} the main conclusion is that the contribution Eq.~(\ref{gammatwophoton}) represents the radiative correction to the one-photon level widths, but not the two-photon transition rate since it can be negative, see~\cite{ourtwoloop,physrep2018}. 

In the presence of BBR, there is an additional line broadening due to the induced transitions. Within the framework of finite temperature QED for bound states~\cite{solovyevarxiv2019}, induced one-photon width $ \Gamma_{a}^{1\gamma,\mathrm{BBR}} $ is given by the imaginary part of thermal one-loop electron self-energy~\cite{solovyev2015}. In the nonrelativistic limit, the result of such evaluation leads to the well-known quantum mechanical expression~\cite{farley}, which is
\begin{eqnarray}
\label{gammaqm}
\Gamma_{a}^{1\gamma,\mathrm{BBR}}=\frac{4e^2}{3}\frac{1}{2l_{a}+1}\sum\limits_{b}\sum\limits_{m_am_b}|\omega_{ab}|^3|\langle b | \textbf{r}| a \rangle |^2n_{\beta}(|\omega_{ab}|).
\end{eqnarray}
Here, summation over states $ b $ extends to the entire spectrum of Schr\"odinger equation,  including upper, lower and continuum states. 

In~\cite{solovyev2015} it was shown that the expression (\ref{gammaqm}) arises in the framework of the QED theory in the approximation of zero level widths. In turn, taking into account the finite lifetimes of atomic levels, the more general expression for the one-photon BBR-induced line broadening can be obtained:
\begin{eqnarray}
\label{gammaqed}
\Gamma_{a}^{1\gamma,\mathrm{BBR-QED}}=
\frac{2e^2}{3\pi}\frac{1}{2l_{a}+1}\sum\limits_{b}\sum\limits_{m_am_b}
|\langle a |\textbf{r}| b \rangle|^2
\int\limits_{0}^{\infty}d\omega n_\beta(\omega) \omega^3
\left[\frac{\Gamma_{ba}}{(\tilde{\omega}_{ba}+\omega)^2 + \frac{1}{4}\Gamma_{ba}^2} + \frac{\Gamma_{ba}}{(\tilde{\omega}_{ba}-\omega)^2 + \frac{1}{4}\Gamma_{ba}^2}\right]
,
\end{eqnarray}
where $ \Gamma_{ba}=\Gamma_{b}+\Gamma_{a} $ is the sum of natural widths of states $ b $ and $ a $. It is easy to see that in the limit $ \Gamma_{ba}\rightarrow 0 $, Eq.~(\ref{gammaqed}) turns to Eq.~(\ref{gammaqm}). The comparison of $ \Gamma_{2s}^{1\gamma,\mathrm{BBR}} $ and $ \Gamma_{2s}^{1\gamma,\mathrm{BBR-QED}} $ for the $ 2s $ state in H and He$^+$ atoms is presented in Table~\ref{tab_comparison}. In particular, it can be found that the accounting for finite lifetimes plays important role for the broadening of $ 2s $ state at low temperatures and is negligible at high temperatures, see also~\cite{zalialiutdinov2019}. 
\begin{table}
\caption{BBR-induced level widths $ \Gamma^{1\gamma,\mathrm{BBR}}_{2s} $ (see Eq.~(\ref{gammaqm})) and $ \Gamma^{1\gamma,\mathrm{BBR-QED}}_{2s} $ (see Eq.~(\ref{gammaqed})) in $ s^{-1} $ at different temperatures $ T $ (in Kelvin) for H and He$^+$ atoms. The Lamb shift $ 2p-2s $ is taken into account \cite{lambH}. The values marked with an asterisks $ ^{*} $ and $ ^{**} $ are taken from \cite{solovyev2015,farley} and \cite{jentbbr} respectively.}
\begin{tabular}{l l c c c c c c c}
\hline
\hline
atom & $T$ & 77 & 300 & 1000 & 3000  & 5000 & 10$^4 $ \\
\hline
H & $ \Gamma_{2s}^{1\gamma,\mathrm{BBR}}     $ & $ 3.653\times 10^{-6} $ & $ 1.423\times 10^{-5} $ & $ 2.023\times 10^{-2} $ & $ 4.701\times 10^{4} $ & $ 9.671\times 10^{5} $ & $ 1.248\times 10^{7} $\\
& & & $ 1.42\times 10^{-5}~^{*} $ & & $ 4.706\times 10^{4}~^{**}$ &  & & \\
  & $ \Gamma_{2s}^{1\gamma,\mathrm{BBR-QED}} $ & $ 2.766\times 10^{-4} $ & $ 4.159\times 10^{-3} $ & $ 6.633\times 10^{-2} $ & $ 4.701\times 10^{4} $ & $ 9.671\times 10^{5} $ & $ 1.248\times 10^{7} $\\
  & $ \Gamma_{3s}^{1\gamma,\mathrm{BBR}}         $ & $ 2.328\times 10^{-6} $ & $ 8.035\times 10^{-5} $ & $ 4.346\times 10^3 $ & $ 9.903\times 10^5 $ & $ 4.042\times 10^6 $ & $ 1.772\times 10^7 $ \\
  & & & $ 7.97\times 10^{-5}~^{*} $ & & & & & \\
  & $ \Gamma_{3s}^{1\gamma,\mathrm{BBR-QED}}     $ & $ 5.148\times 10^{-4} $ & $ 7.865\times 10^{-3}$ & $ 4.346\times 10^3  $ & $ 9.903\times 10^5 $ & $ 4.042\times 10^6 $ & $ 1.772\times 10^7 $ &\\
\hline
\hline
He$^+$ & $ \Gamma_{2s}^{1\gamma,\mathrm{BBR}}     $ & $ 1.599\times 10^{-4} $ & $ 6.249\times 10^{-4} $ & $ 2.085\times 10^{-3} $ & $ 6.472\times 10^{-3} $ & $ 25.97 $ & $ 1.712\times 10^{5} $\\
 & $ \Gamma_{2s}^{1\gamma,\mathrm{BBR-QED}} $ & $ 1.251\times 10^{-3} $ & $ 1.720\times 10^{-2} $ & $ 1.863\times 10^{-1} $ & $ 1.665     $ & $ 30.58 $ & $ 1.712\times 10^{5}  $\\
   & $ \Gamma_{3s}^{1\gamma,\mathrm{BBR}}         $ & $ 8.596\times 10^{-5} $ & $ 3.352\times 10^{-4} $ & $ 1.125\times 10^{-3} $ & $ 5.318\times 10^3 $ & $ 3.296\times 10^5 $ & $ 8.621\times 10^6 $ \\
  & $ \Gamma_{3s}^{1\gamma,\mathrm{BBR-QED}}      $ & $ 4.053\times 10^{-3} $ & $ 6.057\times 10^{-2} $ & $ 6.708\times 10^{-1} $ & $ 5.324\times 10^3 $ & $ 3.296\times 10^5 $ & $ 8.621\times 10^6 $ &\\
\hline
\hline
\end{tabular}
\label{tab_comparison}
\end{table}

Finally, the total level widths of atomic state $ a $ in the presence of BBR field can be written as
\begin{eqnarray}
\label{totalwidths}
\Gamma_{a}^{\mathrm{tot}}=\Gamma_{a}+\Gamma_{a}^{\mathrm{BBR}}
,
\end{eqnarray}
where $ \Gamma_{a} $ is the "zero-temperature" contribution Eq. (\ref{zerot}), and 
\begin{eqnarray}
\label{finitet}
\Gamma_{a}^{\mathrm{BBR}}=\Gamma_{a}^{1\gamma,\mathrm{BBR}}+\Gamma_{a}^{2\gamma,\mathrm{BBR}}+\dots
\end{eqnarray}
represents the contributions (one-photon, two-photon and etc, respectively) corresponding to the BBR-induced level broadening.

To describe the thermal induced two-photon contribution to the level broadening, $ \Gamma_{a}^{2\gamma,\mathrm{BBR}} $, that arises in an atom under the influence of blackbody radiation, we employ the formalism of QED theory at finite temperatures, see for example~\cite{solovyevarxiv2019} and references therein. To take into account the "heat bath" influence on an atom in the framework of this formalism, it is sufficient to consider sequentially the insertions of thermal part of the photon propagator instead of ordinary one in the Feynman graphs that are Figs.~\ref{fig3}-\ref{fig8} in our case. As a result, the imaginary part of two-loop SE corrections, Figs.~\ref{fig3}-\ref{fig8}, in addition to the two-photon processes, contains also the thermal radiative corrections to the one-photon transitions~\cite{saperstein}. Recently, these contributions were considered in the work~\cite{onephoton} and we omit their description here. In turn, the induced two-photon decay widths result from the integration over the pole in the middle electron propagator of the irreducible Feynman diagrams depicted in Figs.~\ref{fig3}-\ref{fig8}~\cite{saperstein,sucher}. 

According to~\cite{labbook}, the corrections $ \Delta E_a $ to the energy of the state $ a $ for any irreducible graphs can be obtained using the relations
\begin{eqnarray}
\label{energyshift}
\Delta E_{a}=\langle a|\hat{U}_{\mathrm{irr}}|a\rangle
,
\end{eqnarray}
where $ \langle a'|\hat{U}_{\mathrm{irr}}|a\rangle $ is the matrix element of amplitude of S-matrix 
\begin{eqnarray}
\label{amplitude}
\langle a'|\hat{S}|a\rangle = -2\pi \mathrm{i} \delta (E_{a'}-E_a)\langle a'|\hat{U}_{\mathrm{irr}}|a\rangle
.
\end{eqnarray}
Then the two-photon radiative correction to the level width is defined by the imaginary part of the second order SE level shift $ \Delta E^{(2)}_{a} $, which is given by the sum of Feynman diagrams Figs.~\ref{fig3}-\ref{fig8}. The result can be presented as
\begin{eqnarray}
\label{gamma}
\Gamma^{2\gamma}_{a}=-2\textrm{Im}\Delta E^{(2)}_{a}
.
\end{eqnarray}

It should be noted here again that the expression (\ref{gamma}) reproduces the two-photon level width in the absence of cascade processes and is merely the radiative correction in general case. In further we give the step by step description of each diagram in Figs.~\ref{fig3}-\ref{fig8} within the $S$-matrix formalism. For clarity, the replacements of "ordinary" loop by the thermal one (one and two times) are denoted by the bold wavy line in the each irreducible graph.

\subsection{Ordinary loop inside thermal loop}

\begin{figure}[hbtp]
\caption{Ordinary loop inside thermal loop Feynman diagram. Bold wavy line denotes thermal photon propagator. A double line is the electron propagator in the Furry picture.}
\centering
\includegraphics[scale=0.6]{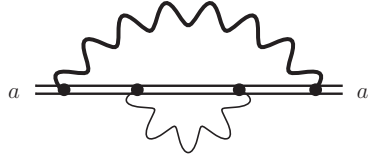}
\label{fig3}
\end{figure}

We start from the consideration of diagram Fig.~\ref{fig3}.
 The corresponding $S$-matrix element is
\begin{eqnarray}
\label{1}
\hat{S}^{(4)\;\mathrm{Fig.~3}}_{aa}= (-\mathrm{i}e)^4\int d^4x_1d^4x_2d^4x_3d^4x_4\overline{\psi}_a(x_4)\gamma_{\mu_4}S(x_4x_3)\gamma_{\mu_3}S(x_3x_2)\gamma_{\mu_2}S(x_2x_1)
\gamma_{\mu_1}\psi_a(x_1)
\\\nonumber\times
 D^{\beta}_{\mu_4\mu_1}(x_4x_1)D_{\mu_3\mu_2}(x_3x_2)
 ,
\end{eqnarray}
where $D_{\mu_i\mu_j}(x_ix_j)$ is the "ordinary" photon propagator. In the Feynman gauge it is
\begin{eqnarray}
\label{2}
D_{\mu_i\mu_j}(x_ix_j)=\frac{1}{2\pi \mathrm{i}}\frac{g_{\mu_i\mu_j}}{r_{ij}}
 \int\limits_{-\infty}^{\infty}d\omega_1 e^{\mathrm{i}\omega(t_i-t_j)+\mathrm{i}|\omega|r_{ij}}.
\end{eqnarray}
$ D^{\beta}_{\mu_i\mu_j}(x_ix_j) $ corresponds to the thermal part of photon propagator~\cite{solovyev2015,solovyevarxiv2019}, which can be reduced to
\begin{eqnarray}
\label{3}
D^{\beta}_{\mu_i\mu_j}(x_ix_j)=- \frac{g_{\mu_i\mu_j}}{\pi r_{12}}\int\limits^{\infty}_{-\infty}d\omega n_{\beta}(|\omega|) \mathrm{sin}(|\omega|r_{ij})e^{-\mathrm{i}\omega(t_i-t_j)}
.
\end{eqnarray}
Then, performing integration over time variables in Eq.~(\ref{1}) and using Eq.~(\ref{amplitude}), the amplitude of the process Fig.~\ref{fig3} is given by
\begin{eqnarray}
\label{4}
U_{a}^{\mathrm{Fig.~3} }=
\sum\limits_{n_1n_2n_3}
\frac{-\mathrm{i}e^4}{2\pi^2}
\int\limits_{-\infty}^{\infty}\int\limits_{-\infty}^{\infty}d\omega_1 d\omega_2 n_{\beta}
(|\omega_1|)
\left[
\frac{1-\bm{\alpha}_{1}\bm{\alpha}_{4}}{r_{14}}\mathrm{sin}(|\omega_1|r_{14})
\right]_{an_1n_3a}
\left[
\frac{1-\bm{\alpha}_{2}\bm{\alpha}_{3}}{r_{23}}e^{\mathrm{i}|\omega_2|r_{23}}
\right]
_{n_2n_3n_1n_2}
\\\nonumber
\times
\frac{1}{\left(E_{n_3}(1-\mathrm{i}0)-E_{a}+\omega_1  \right) \left(E_{n_2}(1-\mathrm{i}0)-E_{a}+\omega_1+\omega_2  \right)  \left(E_{n_1}(1-\mathrm{i}0)-E_{a}+\omega_1  \right)}
.
\end{eqnarray}
The matrix elements $ \left[F(12)\right]_{abcd} $ should be understood as $ \left[F(12)\right]_{a(1)b(2)c(1)d(2)} $, where indexes 1, 2 denote the variables and $ \bm{\alpha}_i $ are the Dirac matrices.

As it was mentioned above the two-photon contribution to the level widths is defined by the pole contribution in the middle electron propagator of Eq.~(\ref{4}). The corresponding integration over $ \omega_2 $ in Eq.~(\ref{4}) can be performed with the use of Cauchy theorem. Repeating the procedure described in~\cite{labbook}, one can obtain
\begin{eqnarray}
\label{labINT}
\int\limits_{-\infty}^{\infty}d\omega_2\frac{e^{\mathrm{i}|\omega_2|r_{23}}}{E_{n_2}(1-\mathrm{i}0)-E_{a}+\omega_1+\omega_2  }
=\frac{\pi\;\mathrm{i}}{2}\;\left(1+\frac{E_{n_2}}{|E_{n_2}|}\right)\left(1-\frac{\omega_{n_2a}+\omega_1}{|\omega_{n_2a}+\omega_1|}\right)
e^{\mathrm{i}|\omega_{n_2a}+\omega_1|r_{23}}
\\\nonumber
 +2\,\mathrm{i}\frac{\omega_{n_2a}+\omega_1}{|\omega_{n_2a}+\omega_1|}
 \left[\mathrm{ci}\left(|\omega_{n_2a}+\omega_1|r_{ij}\right)\,\sin\left(|\omega_{n_2a}+\omega_1|r_{23}\right)\right. 
 -\left.\mathrm{si}\left(|\omega_{n_2a}+\omega_1|r_{ij}\right)\,\cos\left(|\omega_{n_2a}+\omega_1|r_{23}\right)\right],
\end{eqnarray}
where $ \omega_{n_2a}=E_{n_2}-E_{a} $. Then substituting Eq.~(\ref{4}) into Eq.~(\ref{gamma}), using Eq.~(\ref{labINT}) and taking into account the explicit insertion of the remaining modulus sign, we arrive at
\begin{eqnarray}
\label{5}
\Gamma_{a}^{\mathrm{Fig.~3} }=
\frac{2e^4}{\pi}\sum\limits_{n_1n_2n_3}
\left(
\mathrm{Re}\int\limits_{0}^{\infty}d\omega_1n_{\beta}(\omega_1)
\left[\frac{1-\bm{\alpha_1}\bm{\alpha_4}}{r_{14}}
\mathrm{sin(\omega_1r_{14})}
\right]_{an_2n_3a}
\left[
\frac{1-\bm{\alpha_2}\bm{\alpha_3}}{r_{23}}
\mathrm{sin}(|\omega_{n_2a}-\omega_1|r_{23})
\right]_{n_2n_3n_1n_2}
\right.
\\\nonumber
\times
\frac{1}{(E_{n_3}(1-\mathrm{i}0)-E_{a}-\omega_1 )(E_{n_1}(1-\mathrm{i}0)-E_{a}-\omega_1 )}
\left\lbrace  \frac{\pi}{2}\;\left(1+\frac{E_{n_2}}{|E_{n_2}|}\right)\left(1-\frac{\omega_{n_2a}-\omega_1}{|\omega_{n_2a}-\omega_1|}\right)
\right\rbrace
\\\nonumber
+
\mathrm{Re}\int\limits_{0}^{\infty}d\omega_1n_{\beta}(\omega_1)
\left[\frac{1-\bm{\alpha_1}\bm{\alpha_4}}{r_{14}}
\mathrm{sin(\omega_1r_{14})}
\right]_{an_2n_3a}
\left[
\frac{1-\bm{\alpha_2}\bm{\alpha_3}}{r_{23}}
\mathrm{sin}(|\omega_{n_2a}+\omega_1|r_{23})
\right]_{n_2n_3n_1n_2}
\\\nonumber
\times
\left.
\frac{1}{(E_{n_3}(1-\mathrm{i}0)-E_{a}+\omega_1)(E_{n_1}(1-\mathrm{i}0)-E_{a}+\omega_1 ) }
\left\lbrace  \frac{\pi}{2}\;\left(1+\frac{E_{n_2}}{|E_{n_2}|}\right)\left(1-\frac{\omega_{n_2a}+\omega_1}{|\omega_{n_2a}+\omega_1|}\right)
\right\rbrace
\right)
.
\end{eqnarray}

For further evaluation, the two cases should be considered separately: $ \omega_{n_2a}>0 $ and $ \omega_{n_2a}<0 $. First, $ \omega_{n_2a}>0 $, then the second integral in Eq.~(\ref{5}) vanishes, because $ \omega_{n_2a}+\omega_1 $ is always positive, and the integration interval runs through the positive half-axis. Then, for the positive energies $ E_{n_2}>0 $, Eq.~(\ref{5}) reduces to
\begin{eqnarray}
\label{6}
\Gamma_{a}^{\mathrm{Fig.~3} }
=
\frac{2e^4}{\pi}
\sum\limits_{n_1n_2n_3}
\mathrm{Re}\int\limits_{|\omega_{n_2a}|}^{\infty}d\omega n_{\beta}(\omega)
\left[\frac{1-\bm{\alpha_1}\bm{\alpha_4}}{r_{14}}
\mathrm{sin(\omega r_{14})}
\right]_{an_1n_3a}
\left[
\frac{1-\bm{\alpha_2}\bm{\alpha_3}}{r_{23}}
\mathrm{sin}((\omega-|\omega_{n_2a} |)r_{23})
\right]_{n_2n_3n_1n_2}
\\\nonumber\times
\frac{1}{(E_{n_3}(1-\mathrm{i}0)-E_{a}-\omega)(E_{n_1}(1-\mathrm{i}0)-E_{a}-\omega ) }  
.
\end{eqnarray}
For the second case, $ \omega_{n_2a}<0 $, and positive energies $ E_{n_2}>0 $, Eq.~(\ref{5}) transforms to
\begin{eqnarray}
\label{7}
\Gamma_{a}^{\mathrm{Fig.~3} }=\frac{2e^4}{\pi}
\sum\limits_{n_1n_2n_3}
\mathrm{Re}\int\limits_{0}^{\infty}d\omega n_{\beta}(\omega)
\left[
\frac{1-\bm{\alpha_1}\bm{\alpha_4}}{r_{14}}
\mathrm{sin}(\omega r_{14})
\right]_{an_1n_3a}
\left[
\frac{1-\bm{\alpha_2}\bm{\alpha_3}}{r_{23}}
\mathrm{sin}((|\omega_{n_2a}|+\omega)r_{23}) 
\right]_{n_2n_3n_1n_2}
\\\nonumber
\times
\frac{1}{(E_{n_3}(1-\mathrm{i}0)-E_{a}-\omega)(E_{n_1}(1-\mathrm{i}0)-E_{a}-\omega ) }
\\\nonumber
+
\frac{2e^4}{\pi}
\sum\limits_{n_1n_2n_3}
\mathrm{Re}\int\limits_{0}^{|\omega_{n_2a}|}d\omega n_{\beta}(\omega)
\left[
\frac{1-\bm{\alpha_1}\bm{\alpha_4}}{r_{14}}
\mathrm{sin}(\omega r_{14})
\right]
_{an_1n_3a}
\left[
\frac{1-\bm{\alpha_2}\bm{\alpha_3}}{r_{23}}
\mathrm{sin}((|\omega_{n_2a}|-\omega)r_{23}) 
\right]_{n_2n_3n_1n_2}
\\\nonumber
\times
\frac{1}{(E_{n_3}(1-\mathrm{i}0)-E_{a}+\omega)(E_{n_1}(1-\mathrm{i}0)-E_{a}+\omega) }
.
\end{eqnarray}

It will be shown below, see section~\ref{sectionC}, that Eq.~(\ref{6}) and the first term in  Eq.~(\ref{7}) represent a part of thermal correction to the level broadening associated with the Stokes and anti-Stokes Raman scattering rate of thermal photons on virtual states: $ a + \gamma_{\mathrm{T}} \rightarrow n_2 + \gamma $. The second term in Eq.~(\ref{7}) gives correction to the induced two-photon emission.

\subsection{Thermal loop inside ordinary loop}

\begin{figure}[hbtp]
\caption{Thermal loop inside ordinary loop Feynman diagram. All the notations are the same as in Fig.~\ref{fig3}.}
\centering
\includegraphics[scale=0.6]{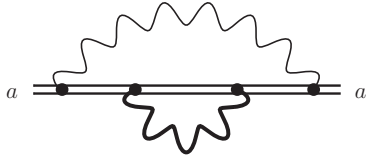}
\label{fig4}
\end{figure}

The evaluation of the digram in Fig.~\ref{fig4} repeats the calculations performed in previous subsection, its $S$-matrix element reads
\begin{eqnarray}
\label{8}
\hat{S}^{(4)\;\mathrm{Fig.~4}}_{aa}= (-\mathrm{i}e)^4\int d^4x_1d^4x_2d^4x_3d^4x_4\overline{\psi}_a(x_4)\gamma_{\mu_4}S(x_4x_3)\gamma_{\mu_3}S(x_3x_2)\gamma_{\mu_2}S(x_2x_1)
\gamma_{\mu_1}\psi_a(x_1)
\\\nonumber\times
 D_{\mu_4\mu_1}(x_4x_1)D^{\beta}_{\mu_3\mu_2}(x_3x_2)
 .
\end{eqnarray}

Integration over the time variables in Eq.~(\ref{8}) leads to the following amplitude:
\begin{eqnarray}
\label{9}
U_{a}^{\mathrm{Fig.~4} }=
\frac{-\mathrm{i}e^4}{2\pi^2}
\sum\limits_{n_1n_2n_3}
\int\limits_{-\infty}^{\infty}\int\limits_{-\infty}^{\infty}d\omega_1 d\omega_2 n_{\beta}
(|\omega_1|)
\left[
\frac{1-\bm{\alpha}_{1}\bm{\alpha}_{4}}{r_{23}}\mathrm{sin}(|\omega_1|r_{23})
\right]_{an_1n_3a}
\left[
\frac{1-\bm{\alpha}_{2}\bm{\alpha}_{3}}{r_{14}}e^{\mathrm{i}|\omega_2|r_{14}}
\right]_{n_2n_3n_1n_2}
\\\nonumber
\times
\frac{1}{\left(E_{n_3}(1-\mathrm{i}0)-E_{a}+\omega_2  \right) \left(E_{n_2}(1-\mathrm{i}0)-E_{a}+\omega_1+\omega_2  \right)  \left(E_{n_1}(1-\mathrm{i}0)-E_{a}+\omega_2  \right)}
.
\end{eqnarray}
Performing integration over $ \omega_2 $ in Eq.~(\ref{9}) with the use of Eq.~(\ref{labINT}), for $ \omega_{n_2a}>0 $ we arrive at
\begin{eqnarray}
\label{10}
\Gamma_{a}^{\mathrm{Fig.~4}}
=
\frac{2e^4}{\pi}
\sum\limits_{n_1n_2n_3}
\mathrm{Re}\int\limits_{|\omega_{n_2a}|}^{\infty}d\omega n_{\beta}(\omega)
\left[
\frac{1-\bm{\alpha_2}\bm{\alpha_3}}{r_{23}}
\mathrm{sin}(\omega r_{23})
\right]_{an_1n_3a}
\left[
\frac{1-\bm{\alpha_1}\bm{\alpha_4}}{r_{14}}
\mathrm{sin}((\omega -|\omega_{n_2a}| ) r_{14})
\right]_{n_2n_3n_1n_2}
\\\nonumber\times
\frac{1}{(E_{n_3}(1-\mathrm{i}0)-E_{n_2}+\omega)(E_{n_1}(1-\mathrm{i}0)-E_{n_2}+\omega ) }  
,
\end{eqnarray}
and for $ \omega_{n_2a}<0 $ we have
\begin{eqnarray}
\label{11}
\Gamma_{a}^{\mathrm{Fig.~4}}
=
\frac{2e^4}{\pi}
\sum\limits_{n_1n_2n_3}
\mathrm{Re}
\int\limits_{0}^{\infty}d\omega n_{\beta}(\omega)
\left[
\frac{1-\bm{\alpha_2}\bm{\alpha_3}}{r_{23}}
\mathrm{sin}(\omega r_{23})
\right]_{an_1n_3a}
\left[
\frac{1-\bm{\alpha_1}\bm{\alpha_4}}{r_{14}}
\mathrm{sin}((|\omega_{n_2a}|+\omega)r_{14}) 
\right]_{n_2n_3n_1n_2}
\\\nonumber
\times
\frac{1}{(E_{n_3}(1-\mathrm{i}0)-E_{n_2}+\omega)(E_{n_1}(1-\mathrm{i}0)-E_{n_2}+\omega )}
\\\nonumber
+
\frac{2e^4}{\pi}
\sum\limits_{n_1n_2n_3}
\mathrm{Re}\int\limits_{0}^{|\omega_{n_2a}|}d\omega n_{\beta}(\omega)
\left[
\frac{1-\bm{\alpha_2}\bm{\alpha_3}}{r_{23}}
\mathrm{sin}(\omega r_{23})
\right]_{an_1n_3a}
\left[
\frac{1-\bm{\alpha_1}\bm{\alpha_4}}{r_{14}}
\mathrm{sin}((|\omega_{n_2a}|-\omega)r_{14}) 
\right]_{n_2n_3n_1n_2}
\\\nonumber
\times
\frac{1}{(E_{n_3}(1-\mathrm{i}0)-E_{n_2}-\omega)(E_{n_1}(1-\mathrm{i}0)-E_{n_2}-\omega)}.
\end{eqnarray}

\subsection{Thermal loop over ordinary loop}

\begin{figure}[hbtp]
\caption{Thermal loop over ordinary loop.}
\centering
\includegraphics[scale=0.6]{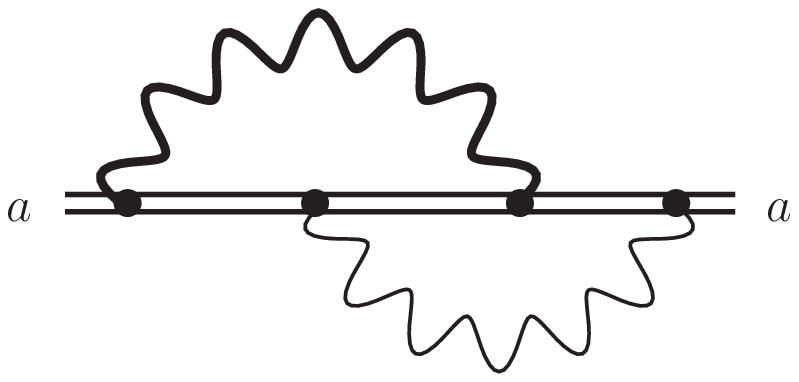}
\label{fig5}
\includegraphics[scale=0.6]{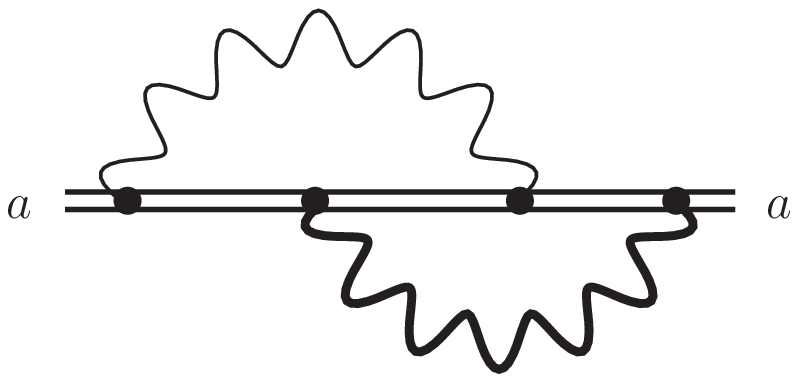}
\label{fig6}
\end{figure}

Contribution corresponding to the Feynman graph depicted in Figs.~\ref{fig5} is given by the following $S$-matrix element:
\begin{eqnarray}
\label{12}
\hat{S}^{(4)\;\mathrm{Fig.~}}_{aa}= (-\mathrm{i}e)^4\int d^4x_1d^4x_2d^4x_3d^4x_4\overline{\psi}_a(x_4)\gamma_{\mu_4}S(x_4x_3)\gamma_{\mu_3}S(x_3x_2)\gamma_{\mu_2}S(x_2x_1)
\gamma_{\mu_1}\psi_a(x_1)
\\\nonumber\times
 D_{\mu_4\mu_2}(x_4x_2)D^{\beta}_{\mu_3\mu_1}(x_3x_1)
 .
\end{eqnarray}

Integration over the time variables in Eq.~(\ref{12}) leads to the expression for amplitude
\begin{eqnarray}
\label{13}
U_{a}^{\mathrm{Fig.~5}}=
\frac{-\mathrm{i}e^4}{2\pi^2}
\sum\limits_{n_1n_2n_3}
\int\limits_{-\infty}^{\infty}\int\limits_{-\infty}^{\infty}d\omega_1 d\omega_2 n_{\beta}
(|\omega_1|)
\left[
\frac{1-\bm{\alpha}_{3}\bm{\alpha}_{1}}{r_{31}}\mathrm{sin}(|\omega_1|r_{31})
\right]_{an_2n_3n_1}
\left[
\frac{1-\bm{\alpha}_{4}\bm{\alpha}_{2}}{r_{42}}e^{\mathrm{i}|\omega_2|r_{42}}
\right]_{n_3n_1n_2a}
\\\nonumber
\times
\frac{1}{\left(E_{n_3}(1-\mathrm{i}0)-E_{a}+\omega_1  \right) \left(E_{n_2}(1-\mathrm{i}0)-E_{a}+\omega_1+\omega_2  \right)  \left(E_{n_1}(1-\mathrm{i}0)-E_{a}+\omega_2  \right)}
.
\end{eqnarray}
Performing integration over $ \omega_2 $ in Eq.~(\ref{13}), using Eq.~(\ref{labINT}) and repeating steps described above, we obtain
\begin{eqnarray}
\label{14}
\Gamma_{a}^{\mathrm{Fig.~5}}
=
\frac{2e^4}{\pi}
\sum\limits_{n_1n_2n_3}
\mathrm{Re}\int\limits_{|\omega_{n_2a}|}^{\infty}d\omega 
n_{\beta}(\omega)
\left[\frac{1-\bm{\alpha_3}\bm{\alpha_1}}{r_{31}}
\mathrm{sin}(\omega r_{31})
\right]_{an_2n_3n_1}
\left[
\frac{1-\bm{\alpha_4}\bm{\alpha_2}}{r_{42}}
\mathrm{sin}((\omega-|\omega_{n_2a}| ) r_{42})
\right]_{n_3n_1n_2a}
\\\nonumber\times
\frac{1}{(E_{n_3}(1-\mathrm{i}0)-E_{a}-\omega)(E_{n_1}(1-\mathrm{i}0)-E_{n_2}+\omega ) }  
,
\end{eqnarray}
where $ \omega_{n_2a}>0 $, and for $ \omega_{n_2a}<0 $ we have
\begin{eqnarray}
\label{15}
\Gamma_{a}^{\mathrm{Fig5}}
=
\frac{2e^4}{\pi}
\sum\limits_{n_1n_2n_3}
\mathrm{Re}
\int\limits_{0}^{\infty}d\omega n_{\beta}(\omega)
\left[
\frac{1-\bm{\alpha_3}\bm{\alpha_1}}{r_{31}}
\mathrm{sin}(\omega r_{31})
\right]_{an_2n_3n_1}
\left[
\frac{1-\bm{\alpha_4}\bm{\alpha_2}}{r_{42}}
\mathrm{sin}((|\omega_{n_2a}|+\omega)r_{42}) 
\right]_{n_3n_1n_2a}
\\\nonumber
\times
\frac{1}{(E_{n_3}(1-\mathrm{i}0)-E_{a}-\omega)(E_{n_1}(1-\mathrm{i}0)-E_{n_2}+\omega ) }
\\\nonumber
+
\frac{2e^4}{\pi}
\sum\limits_{n_1n_2n_3}
\mathrm{Re}\int\limits_{0}^{|\omega_{n_2a}|}d\omega n_{\beta}(\omega)
\left[
\frac{1-\bm{\alpha_3}\bm{\alpha_1}}{r_{31}}
\mathrm{sin}(\omega r_{31})
\right]_{an_2n_3n_1}
\left[
\frac{1-\bm{\alpha_4}\bm{\alpha_2}}{r_{42}}
\mathrm{sin}((|\omega_{n_2a}|+\omega)r_{42}) 
\right]_{n_3n_1n_2a}
\\\nonumber
\times
\frac{1}{(E_{n_3}(1-\mathrm{i}0)-E_{a}+\omega)(E_{n_1}(1-\mathrm{i}0)-E_{n_2}-\omega) }.
\end{eqnarray}

\subsection{Thermal loop inside thermal loop}

Contribution of the Feynman graph depicted in Fig.~\ref{fig6} corresponds to the $S$-matrix element:
\begin{eqnarray}
\label{16}
\hat{S}^{(4)\;\mathrm{Fig.~6}}_{aa}= (-\mathrm{i}e)^4\int d^4x_1d^4x_2d^4x_3d^4x_4\overline{\psi}_a(x_1)\gamma_{\mu_4}S(x_1x_2)\gamma_{\mu_3}S(x_2x_3)\gamma_{\mu_2}S(x_3x_4)
\gamma_{\mu_1}\psi_a(x_4)
\\\nonumber\times
D^{\beta}_{\mu_1\mu_4}(x_1x_4)D^{\beta}_{\mu_2\mu_3}(x_2x_3)
.
\end{eqnarray}
\begin{figure}[hbtp]
\caption{Thermal loop inside thermal loop.}
\centering
\includegraphics[scale=0.6]{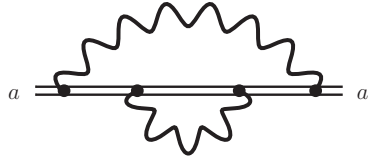}
\label{fig7}
\end{figure}
Integration over the time variables in Eq.~(\ref{16}) leads to
\begin{gather}
\label{17}
U_{a}^{\mathrm{Fig.~6} }=
-\frac{e^4}{\pi^2}
\sum\limits_{n_1n_2n_3}
\int\limits_{-\infty}^{\infty}\int\limits_{-\infty}^{\infty}d\omega_1 d\omega_2 
n_{\beta}(|\omega_1|)
n_{\beta}(|\omega_2|)
\\\nonumber\times
\left[
\frac{1-\bm{\alpha}_{1}\bm{\alpha}_{4}}{r_{14}}\mathrm{sin}(|\omega_1|r_{14})
\right]_{an_1n_3a}
\left[
\frac{1-\bm{\alpha}_{2}\bm{\alpha}_{3}}{r_{23}}\mathrm{sin}(|\omega_2|r_{23})
\right]_{n_2n_3n_1n_2}
\\\nonumber
\times
\frac{1}{\left(E_{n_3}(1-\mathrm{i}0)-E_{a}+\omega_1  \right) \left(E_{n_2}(1-\mathrm{i}0)-E_{a}+\omega_1+\omega_2  \right)  \left(E_{n_1}(1-\mathrm{i}0)-E_{a}+\omega_1  \right)}
.
\end{gather}

Then integration over $ \omega_2 $ in Eq.~(\ref{17}) can be performed with the use of the Sokhotski-Plemelj theorem
\begin{eqnarray}
\label{18}
\frac{1}{x\pm \mathrm{i}0}=\mathcal{P}\frac{1}{x} \mp \pi\mathrm{i}\;\delta(x)
,
\end{eqnarray}
where $ \mathcal{P} $ means the principal value of the integral. Taking into account Eq. (\ref{18}), the imaginary part is
\begin{eqnarray}
\label{19}
\Gamma_{a}^{\mathrm{Fig.~6} }=
\frac{2e^4}{\pi}
\sum\limits_{n_1n_2n_3}
\int\limits_{-\infty}^{\infty}
\int\limits_{-\infty}^{\infty}
d\omega_1 d\omega_2 
\frac{
n_{\beta}(|\omega_1|)
n_{\beta}(|\omega_2|)
\delta(E_{n_2}-E_{a}+\omega_1+\omega_2)
}
{
\left(E_{n_3}(1-\mathrm{i}0)-E_{a}+\omega_1  \right) 
\left(E_{n_1}(1-\mathrm{i}0)-E_{a}+\omega_1  \right)
}
\\\nonumber
\times
\left[
\frac{1-\bm{\alpha}_{1}\bm{\alpha}_{4}}{r_{14}}\mathrm{sin}(|\omega_1|r_{14})
\right]_{an_1n_3a}
\left[
\frac{1-\bm{\alpha}_{2}\bm{\alpha}_{3}}{r_{23}}\mathrm{sin}(|\omega_2|r_{23})
\right]_{n_2n_3n_1n_2}
\\\nonumber
=
\frac{2e^4}{\pi}
\sum\limits_{n_1n_2n_3}
\left\lbrace
\int\limits_{-\infty}^{\infty}
\right.
d\omega_1 
\int\limits_{0}^{\infty}
d\omega_2 
\frac{
n_{\beta}(|\omega_1|)
n_{\beta}(\omega_2)
\delta(E_{n_2}-E_{a}+\omega_1-\omega_2)
}
{
\left(E_{n_3}(1-\mathrm{i}0)-E_{a}+\omega_1  \right) 
\left(E_{n_1}(1-\mathrm{i}0)-E_{a}+\omega_1  \right)
}
\\\nonumber
\times
\left[
\frac{1-\bm{\alpha}_{1}\bm{\alpha}_{4}}{r_{14}}\mathrm{sin}(|\omega_1|r_{14})
\right]_{an_1n_3a}
\left[
\frac{1-\bm{\alpha}_{2}\bm{\alpha}_{3}}{r_{23}}\mathrm{sin}(\omega_2 r_{23})
\right]_{n_2n_3n_1n_2}
\\\nonumber
+
\int\limits_{-\infty}^{\infty}
d\omega_1 
\int\limits_{0}^{\infty}
d\omega_2 
\frac{
n_{\beta}(|\omega_1|)
n_{\beta}(\omega_2)
\delta(E_{n_2}-E_{a}+\omega_1+\omega_2)
}
{
\left(E_{n_3}(1-\mathrm{i}0)-E_{a}+\omega_1  \right) 
\left(E_{n_1}(1-\mathrm{i}0)-E_{a}+\omega_1  \right)
}
\left[
\frac{1-\bm{\alpha}_{1}\bm{\alpha}_{4}}{r_{14}}\mathrm{sin}(|\omega_1|r_{14})
\right]_{an_1n_3a}
\\\nonumber
\times
\left.
\left[
\frac{1-\bm{\alpha}_{2}\bm{\alpha}_{3}}{r_{23}}\mathrm{sin}(\omega_2 r_{23})
\right]_{n_2n_3n_1n_2}
\right\rbrace
.
\end{eqnarray}

As before, the integration over frequency $ \omega_2 $ in Eq.~(\ref{19}) differs for $ \omega_{n_2a}<0 $ and $ \omega_{n_2a}>0 $. The result for $\omega_{n_2a}>0 $ is
\begin{eqnarray}
\label{20}
\Gamma_{a}^{\mathrm{Fig.~6} }=
\frac{2e^4}{\pi}
\sum\limits_{n_1n_2n_3}
\left\lbrace
\int\limits_{0}^{|\omega_{n_2a}|}d\omega 
\right.
\frac{n_{\beta}(\omega)n_{\beta}(|\omega_{n_2a}|-\omega)}
{(E_{n_3}(1-\mathrm{i}0)-E_{a}-\omega)(E_{n_1}(1-\mathrm{i}0)-E_{a}-\omega)}
\left[
\frac{1-\bm{\alpha}_{1}\bm{\alpha}_{4}}{r_{14}}\mathrm{sin}(\omega r_{14})
\right]_{an_1n_3a}
\\\nonumber
\times
\left[
\frac{1-\bm{\alpha}_{2}\bm{\alpha}_{3}}{r_{23}}\mathrm{sin}((|\omega_{n_2a}|-\omega)r_{23})
\right]_{n_2n_3n_1n_2}
\\\nonumber
+
\int\limits_{|\omega_{n_2a}|}^{\infty}d\omega 
\frac{n_{\beta}(\omega)n_{\beta}(\omega - |\omega_{n_2a}|)}
{(E_{n_3}(1-\mathrm{i}0)-E_{a}-\omega)(E_{n_1}(1-\mathrm{i}0)-E_{a}-\omega)}
\left[
\frac{1-\bm{\alpha}_{1}\bm{\alpha}_{4}}{r_{14}}\mathrm{sin}(\omega r_{14})
\right]_{an_1n_3a}
\\\nonumber
\times
\left[
\frac{1-\bm{\alpha}_{2}\bm{\alpha}_{3}}{r_{23}}\mathrm{sin}((\omega-|\omega_{n_2a}|)r_{23})
\right]_{n_2n_3n_1n_2}
\\\nonumber
+
\int\limits_{0}^{\infty}d\omega 
\frac{n_{\beta}(\omega)n_{\beta}(|\omega_{n_2a}|+\omega)}
{(E_{n_3}(1-\mathrm{i}0)-E_{a}+\omega)(E_{n_1}(1-\mathrm{i}0)-E_{a}+\omega)}
\left[
\frac{1-\bm{\alpha}_{1}\bm{\alpha}_{4}}{r_{14}}\mathrm{sin}(\omega r_{14})
\right]_{an_1n_3a}
\\\nonumber
\times
\left.
\left[
\frac{1-\bm{\alpha}_{2}\bm{\alpha}_{3}}{r_{23}}\mathrm{sin}((|\omega_{n_2a}|+\omega)r_{23})
\right]_{n_2n_3n_1n_2}
\right\rbrace
\end{eqnarray}
 and for $\omega_{n_2a}<0 $
\begin{eqnarray}
\label{21}
\Gamma_{a}^{\mathrm{Fig.~6} }=
\frac{2e^4}{\pi}
\sum\limits_{n_1n_2n_3}
\left\lbrace
\int\limits_{0}^{\infty}d\omega 
\right.
\frac{n_{\beta}(\omega)n_{\beta}(|\omega_{n_2a}|+\omega)}
{(E_{n_3}(1-\mathrm{i}0)-E_{a}-\omega)(E_{n_1}(1-\mathrm{i}0)-E_{a}-\omega)}
\left[
\frac{1-\bm{\alpha}_{1}\bm{\alpha}_{4}}{r_{14}}\mathrm{sin}(\omega r_{14})
\right]_{an_1n_3a}
\\\nonumber
\times
\left[
\frac{1-\bm{\alpha}_{2}\bm{\alpha}_{3}}{r_{23}}\mathrm{sin}((|\omega_{n_2a}|+\omega)r_{23})
\right]_{n_2n_3n_1n_2}
\\\nonumber
+
\int\limits_{0}^{|\omega_{n_2a}|}d\omega 
\frac{n_{\beta}(\omega)n_{\beta}(|\omega_{n_2a}|-\omega)}
{(E_{n_3}(1-\mathrm{i}0)-E_{a}+\omega)(E_{n_1}(1-\mathrm{i}0)-E_{a}+\omega)}
\left[
\frac{1-\bm{\alpha}_{1}\bm{\alpha}_{4}}{r_{14}}\mathrm{sin}(\omega r_{14})
\right]_{an_1n_3a}
\\\nonumber
\times
\left[
\frac{1-\bm{\alpha}_{2}\bm{\alpha}_{3}}{r_{23}}\mathrm{sin}((|\omega_{n_2a}|-\omega)r_{23})
\right]_{n_2n_3n_1n_2}
\\\nonumber
+
\int\limits_{|\omega_{n_2a}|}^{\infty}d\omega 
\frac{n_{\beta}(\omega)n_{\beta}(\omega-|\omega_{n_2a}|)}
{(E_{n_3}(1-\mathrm{i}0)-E_{a}+\omega)(E_{n_1}(1-\mathrm{i}0)-E_{a}+\omega)}
\left[
\frac{1-\bm{\alpha}_{1}\bm{\alpha}_{4}}{r_{14}}\mathrm{sin}(\omega r_{14})
\right]_{an_1n_3a}
\\\nonumber
\times
\left.
\left[
\frac{1-\bm{\alpha}_{2}\bm{\alpha}_{3}}{r_{23}}\mathrm{sin}((\omega-|\omega_{n_2a}|)r_{23})
\right]_{n_2n_3n_1n_2}
\right\rbrace
.
\end{eqnarray}

\subsection{Thermal loop over thermal loop}

\begin{figure}[hbtp]
\caption{Thermal loop over thermal loop.}
\centering
\includegraphics[scale=0.6]{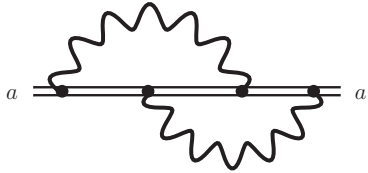}
\label{fig8}
\end{figure}

Contribution corresponding to the Feynman graph depicted in Fig.~\ref{fig8} is given by the $S$-matrix element:
\begin{eqnarray}
\label{22}
\hat{S}^{(4)\;\mathrm{Fig.~7}}_{aa}= (-\mathrm{i}e)^4\int d^4x_1d^4x_2d^4x_3d^4x_4\overline{\psi}_a(x_1)\gamma_{\mu_4}S(x_1x_2)\gamma_{\mu_3}S(x_2x_3)\gamma_{\mu_2}S(x_3x_4)
\gamma_{\mu_1}\psi_a(x_4)
\\\nonumber\times
D^{\beta}_{\mu_1\mu_4}(x_2x_4)D^{\beta}_{\mu_2\mu_3}(x_1x_3).
\end{eqnarray}

Integration over the time variables in Eq.~(\ref{22}) yields
\begin{eqnarray}
U_{a}^{\mathrm{Fig.~7}}=
-\frac{e^4}{\pi^2}
\sum\limits_{n_1n_2n_3}
\int\limits_{-\infty}^{\infty}\int\limits_{-\infty}^{\infty}d\omega_1 d\omega_2 n_{\beta}
(|\omega_1|)
\left[
\frac{1-\bm{\alpha}_{3}\bm{\alpha}_{1}}{r_{31}}\mathrm{sin}(|\omega_1|r_{31})
\right]_{an_2n_3n_1}
\left[
\frac{1-\bm{\alpha}_{4}\bm{\alpha}_{2}}{r_{42}}\mathrm{sin}(|\omega_2|r_{42})
\right]_{n_3n_1n_2a}
\\\nonumber
\times
\frac{1}{\left(E_{n_3}(1-\mathrm{i}0)-E_{a}+\omega_1  \right) \left(E_{n_2}(1-\mathrm{i}0)-E_{a}+\omega_1+\omega_2  \right)  \left(E_{n_1}(1-\mathrm{i}0)-E_{a}+\omega_2  \right)}
\end{eqnarray}
Then, substituting Eq.~(\ref{19}) into Eq.~(\ref{gamma}) and integrating over $ \omega_2 $ according to the Sokhotski-Plemelj theorem, the result for $\omega_{n_2a}>0 $ is
\begin{eqnarray}
\label{20tt}
\Gamma_{a}^{\mathrm{Fig.~7} }=
\frac{2e^4}{\pi}
\left\lbrace
\int\limits_{0}^{|\omega_{n_2a}|}d\omega 
\right.
\frac{n_{\beta}(\omega)n_{\beta}(|\omega_{n_2a}|-\omega)}
{(E_{n_3}(1-\mathrm{i}0)-E_{a}-\omega)(E_{n_1}(1-\mathrm{i}0)-E_{т_2}+\omega)}
\left[
\frac{1-\bm{\alpha}_{3}\bm{\alpha}_{1}}{r_{31}}\mathrm{sin}(\omega r_{31})
\right]_{an_2n_3n_1}
\\\nonumber
\times
\left[
\frac{1-\bm{\alpha}_{4}\bm{\alpha}_{2}}{r_{42}}\mathrm{sin}((|\omega_{n_2a}|-\omega)r_{42})
\right]_{n_3n_1n_2a}
\\\nonumber
+
\int\limits_{|\omega_{n_2a}|}^{\infty}d\omega 
\frac{n_{\beta}(\omega)n_{\beta}(\omega-|\omega_{n_2a}|)}
{(E_{n_3}(1-\mathrm{i}0)-E_{a}-\omega)(E_{n_1}(1-\mathrm{i}0)-E_{n_2}+\omega)}
\left[
\frac{1-\bm{\alpha}_{3}\bm{\alpha}_{1}}{r_{31}}\mathrm{sin}(\omega_1 r_{31})
\right]_{an_2n_3n_1}
\\\nonumber
\left[
\frac{1-\bm{\alpha}_{4}\bm{\alpha}_{2}}{r_{42}}\mathrm{sin}((\omega - |\omega_{n_2a}|)r_{42})
\right]_{n_3n_1n_2a}
\\\nonumber
+
\int\limits_{0}^{\infty}d\omega 
\frac{n_{\beta}(\omega)n_{\beta}(|\omega_{n_2a}|+\omega)}
{(E_{n_3}(1-\mathrm{i}0)-E_{a}+\omega)(E_{n_1}(1-\mathrm{i}0)-E_{n_2}-\omega)}
\left[
\frac{1-\bm{\alpha}_{3}\bm{\alpha}_{1}}{r_{31}}\mathrm{sin}(\omega r_{31})
\right]_{an_2n_3n_1}
\\\nonumber
\times
\left.
\left[
\frac{1-\bm{\alpha}_{4}\bm{\alpha}_{2}}{r_{42}}\mathrm{sin}((|\omega_{n_2a}|+\omega)r_{42})
\right]_{n_3n_1n_2a}
\right\rbrace,
\end{eqnarray}
and for $\omega_{n_2a}<0 $ one can obtain
\begin{eqnarray}
\label{21tt}
\Gamma_{a}^{\mathrm{Fig.~7} }=
\frac{2e^4}{\pi}
\left\lbrace
\int\limits_{0}^{\infty}d\omega 
\right.
\frac{n_{\beta}(\omega)n_{\beta}(|\omega_{n_2a}|+\omega)}
{(E_{n_3}(1-\mathrm{i}0)-E_{a}-\omega)(E_{n_1}(1-\mathrm{i}0)-E_{n_2}+\omega)}
\left[
\frac{1-\bm{\alpha}_{1}\bm{\alpha}_{4}}{r_{14}}\mathrm{sin}(\omega r_{14})
\right]_{an_1n_3a}
\\\nonumber
\times
\left[
\frac{1-\bm{\alpha}_{2}\bm{\alpha}_{3}}{r_{23}}\mathrm{sin}((|\omega_{n_2a}|+\omega)r_{23})
\right]_{n_2n_3n_1n_2}
\\\nonumber
+
\int\limits_{0}^{|\omega_{n_2a}|}d\omega 
\frac{n_{\beta}(\omega)n_{\beta}(|\omega_{n_2a}|-\omega)}
{(E_{n_3}(1-\mathrm{i}0)-E_{a}+\omega)(E_{n_1}(1-\mathrm{i}0)-E_{n_2}-\omega)}
\left[
\frac{1-\bm{\alpha}_{1}\bm{\alpha}_{4}}{r_{14}}\mathrm{sin}(\omega r_{14})
\right]_{an_1n_3a}
\\\nonumber
\times
\left[
\frac{1-\bm{\alpha}_{2}\bm{\alpha}_{3}}{r_{23}}\mathrm{sin}((|\omega_{n_2a}|-\omega)r_{23})
\right]_{n_2n_3n_1n_2}
\\\nonumber
+
\int\limits_{|\omega_{n_2a}|}^{\infty}d\omega 
\frac{n_{\beta}(\omega)n_{\beta}(\omega-|\omega_{n_2a}|)}
{(E_{n_3}(1-\mathrm{i}0)-E_{a}+\omega)(E_{n_1}(1-\mathrm{i}0)-E_{n_2}-\omega)}
\left[
\frac{1-\bm{\alpha}_{1}\bm{\alpha}_{4}}{r_{14}}\mathrm{sin}(\omega r_{14})
\right]_{an_1n_3a}
\\\nonumber
\times
\left.
\left[
\frac{1-\bm{\alpha}_{2}\bm{\alpha}_{3}}{r_{23}}\mathrm{sin}((\omega-|\omega_{n_2a}|)r_{23})
\right]_{n_2n_3n_1n_2}
\right\rbrace.
\end{eqnarray}

\section{Two-loop decay widths at finite temperatures: nonrelativistic limit}
\label{sectionC}

In the present section we collect all the contributions above to form the thermal induced two-photon decay widths. The combination of these contributions should be compared with Eqs.~(\ref{ind}) and (\ref{raman}) for the induced two-photon transitions and Raman scattering of thermal photons. This is reasonable to perform in the nonrelativistic limit, see \cite{labbook}. Then the matrix elements in Eqs.~(\ref{6}), (\ref{7}), (\ref{10}), (\ref{11}), (\ref{14}), (\ref{15}), (\ref{20}), (\ref{21}), (\ref{20tt}) and (\ref{21tt}) can be simplified with
\begin{eqnarray}
\label{nrmatrix}
\left[
\frac{1-\bm{\alpha}_{i}\bm{\alpha}_{j}}{r_{ij}}\mathrm{sin}(\omega r_{ij})
\right]_{a(i)b(j)c(i)d(j)}
\sim \omega  \delta_{ac}\delta_{bd} 
+ 
\left(
- \omega\omega_{ac}\omega_{db}  + \frac{\omega^3}{3} 
\right)
 \langle a |\textbf{r}| c \rangle\langle b |\textbf{r} |d \rangle
 ,
\end{eqnarray}
where $ \delta_{ab} $ is the Kronecker symbol and relation
$
\langle a|\textbf{p}| b \rangle=\mathrm{i}\omega_{ab}\langle a |\textbf{r}| b \rangle
$
was used. 

The first term in the right-hand side of Eq.~(\ref{nrmatrix}) generates infrared divergences of the type $ \int_{0}^{\infty}d\omega n_{\beta}(\omega)/\omega $ for each diagram Figs.~\ref{fig3}-\ref{fig8}. However, in the sum of loop-inside-loop and loop-over-loop diagrams they arise with different signs and, finally, cancel each other. This is the ordinary situation occurring for evaluation of the radiative QED corrections to the emission processes~\cite{shabaevreports}. Recently, the same conclusion was verified for the one-loop self-energy corrections at finite temperature~\cite{onephoton}. It should be noted here, that the regularization of energy shifts suggested in \cite{solovyevarxiv2019} is also fulfilled in this case. All the divergences arising for each Feynman diagram would be canceled by the coincident limit separately, and the final result will be the same as here. The rigorous proof of this repeats the derivations in previous papers and we omit it for brevity.

To evaluate the terms linear on $ \omega $ in the sum of all contributions (corresponding to the second term in Eq.~(\ref{nrmatrix})), the following equality is helpful~\cite{labbook}:
\begin{eqnarray}
\label{gaugetrick}
\omega(\omega_{0}-\omega)\sum\limits_{n}
\omega_{bn}\omega_{an}
\left(
\frac{\langle b|\textbf{r}|n\rangle \langle n |\textbf{r}|a\rangle}{E_{n}(1-\mathrm{i}0)-E_{a}+\omega}+\frac{\langle b|\textbf{r}|n\rangle \langle n |\textbf{r}|a\rangle}{E_{n}(1-\mathrm{i}0)-E_{b}-\omega}
\right)
\\\nonumber
\times
\left(
\frac{\langle b|\textbf{r}|n\rangle^{*}\langle n |\textbf{r}|a\rangle^{*}}{E_{n}(1-\mathrm{i}0)-E_{a}+\omega}+\frac{\langle b|\textbf{r}|n\rangle^{*} \langle n |\textbf{r}|a\rangle^{*}}{E_{n}(1-\mathrm{i}0)-E_{b}-\omega}
\right)
\\\nonumber
=
\omega^3(\omega_{0}-\omega)^3\sum\limits_{n}
\left(
\frac{\langle b|\textbf{r}|n\rangle \langle n |\textbf{r}|a\rangle}{E_{n}(1-\mathrm{i}0)-E_{a}+\omega}+\frac{\langle b|\textbf{r}|n\rangle \langle n |\textbf{r}|a\rangle}{E_{n}(1-\mathrm{i}0)-E_{b}-\omega}
\right)
\\\nonumber
\times
\left(
\frac{\langle b|\textbf{r}|n\rangle^{*}\langle n |\textbf{r}|a\rangle^{*}}{E_{n}(1-\mathrm{i}0)-E_{a}+\omega}+\frac{\langle b|\textbf{r}|n\rangle^{*} \langle n |\textbf{r}|a\rangle^{*}}{E_{n}(1-\mathrm{i}0)-E_{b}-\omega}
\right)
,
\end{eqnarray}
where $ \omega_{0}=E_{a}-E_{b} $. 

Finally, the combination of Eqs.~(\ref{6}), (\ref{7}), (\ref{10}), (\ref{11}), (\ref{14}), (\ref{15}), (\ref{20}), (\ref{21}), (\ref{20tt}) and (\ref{21tt}), as well as  the use of Eq. (\ref{intjent}) and Eqs.~(\ref{nrmatrix}), (\ref{gaugetrick}), the averaging over projections of initial state and summation over projections of final state, results to the thermal two-photon decay width:
\begin{eqnarray}
\label{totalg}
\Gamma_{a}^{2\gamma,\mathrm{BBR}}
=\sum\limits_{b}\Gamma_{ab}^{2\gamma,\mathrm{BBR}}=\sum\limits_{b}
\left(
\Gamma_{ab}^{2\gamma,\mathrm{trans}}+\Gamma_{ab}^{2\gamma,\mathrm{ram}}+\Gamma_{ab}^{2\gamma,\mathrm{int}}
\right)
,
\end{eqnarray}
where for $ b<a $ (i.e. $ \omega_{ba}<0 $)
\begin{gather}
\label{GAm}
\Gamma_{ab}^{2\gamma,\mathrm{trans}}=\frac{4e^4}{9\pi}\frac{1}{2l_{a}+1}
\lim_{\eta\rightarrow 0}\mathrm{Re}
\sum\limits_{m_am_b}\int\limits_{0}^{|\omega_{ba}|}
d\omega\omega^3(|\omega_{ba}|-\omega)^3
\sum\limits_{n}
\left(
\frac{\langle b|\textbf{r}|n\rangle \langle n |\textbf{r}|a\rangle}{E_{n}-E_{a}+\omega+\mathrm{i}\eta}+\frac{\langle b|\textbf{r}|n\rangle \langle n |\textbf{r}|a\rangle}{E_{n}-E_{b}-\omega+\mathrm{i}\eta}
\right)
\\\nonumber
\times
\left(
\frac{\langle b|\textbf{r}|n\rangle^{*}\langle n |\textbf{r}|a\rangle^{*}}{E_{n}-E_{a}+\omega+\mathrm{i}\eta}+\frac{\langle b|\textbf{r}|n\rangle^{*} \langle n |\textbf{r}|a\rangle^{*}}{E_{n}-E_{b}-\omega+\mathrm{i}\eta}
\right)
(n_{\beta}(\omega)+n_{\beta}(|\omega_{ba}|-\omega)+n_{\beta}(\omega)n_{\beta}(|\omega_{ba}|-\omega))
,
\end{gather}
\begin{gather}
\label{GBm}
\Gamma_{ab}^{2\gamma,\mathrm{ram}}=\frac{8e^4}{9\pi}\frac{1}{2l_{a}+1}
\lim_{\eta\rightarrow 0}\mathrm{Re}
\sum\limits_{m_am_b}\int\limits_{0}^{\infty}
d\omega
\omega^3(|\omega_{ba}|+\omega)^3
\sum\limits_{n}
\left(
\frac{\langle b|\textbf{r}|n\rangle \langle n |\textbf{r}|a\rangle}{E_{n}-E_{a}-\omega+\mathrm{i}\eta}+\frac{\langle b|\textbf{r}|n\rangle \langle n |\textbf{r}|a\rangle}{E_{n}-E_{b}+\omega+\mathrm{i}\eta}
\right)
\\\nonumber
\times
\left(
\frac{\langle b|\textbf{r}|n\rangle^{*}\langle n |\textbf{r}|a\rangle^{*}}{E_{n}-E_{a}-\omega+\mathrm{i}\eta}+\frac{\langle b|\textbf{r}|n\rangle^{*} \langle n |\textbf{r}|a\rangle^{*}}{E_{n}-E_{b}+\omega+\mathrm{i}\eta}
\right)
(n_{\beta}(\omega)+n_{\beta}(\omega)n_{\beta}(|\omega_{ba}|+\omega))
,
\end{gather}
\begin{gather}
\label{GCm}
\Gamma_{ab}^{2\gamma,\mathrm{int}}=\frac{8e^4}{9\pi}\frac{1}{2l_{a}+1}
\lim_{\eta\rightarrow 0}\mathrm{Re}
\sum\limits_{m_am_b}\int\limits_{|\omega_{ba}|}^{\infty}
d\omega
\omega^3(\omega-|\omega_{ba}|)^3
\sum\limits_{n}
\left(
\frac{\langle b|\textbf{r}|n\rangle \langle n |\textbf{r}|a\rangle}{E_{n}-E_{a}+\omega+\mathrm{i}\eta}+\frac{\langle b|\textbf{r}|n\rangle \langle n |\textbf{r}|a\rangle}{E_{n}-E_{b}-\omega+\mathrm{i}\eta}
\right)
\\\nonumber
\times
\left(
\frac{\langle b|\textbf{r}|n\rangle^{*}\langle n |\textbf{r}|a\rangle^{*}}{E_{n}-E_{a}+\omega+\mathrm{i}\eta}+\frac{\langle b|\textbf{r}|n\rangle^{*} \langle n |\textbf{r}|a\rangle^{*}}{E_{n}-E_{b}-\omega+\mathrm{i}\eta}
\right)
n_{\beta}(\omega)n_{\beta}(\omega-|\omega_{ba}|),
\end{gather}
and for $ b>a $ (i.e. $ \omega_{ba}>0 $)
\begin{gather}
\label{GAp}
\Gamma_{ab}^{2\gamma,\mathrm{trans}}=\frac{4e^4}{9\pi}\frac{1}{2l_{a}+1}
\lim_{\eta\rightarrow 0}\mathrm{Re}
\sum\limits_{m_am_b}\int\limits_{0}^{|\omega_{ba}|}
d\omega\omega^3(|\omega_{ba}|-\omega)^3
\sum\limits_{n}
\left(
\frac{\langle b|\textbf{r}|n\rangle \langle n |\textbf{r}|a\rangle}{E_{n}-E_{a}-\omega+\mathrm{i}\eta}+\frac{\langle b|\textbf{r}|n\rangle \langle n |\textbf{r}|a\rangle}{E_{n}-E_{b}+\omega+\mathrm{i}\eta}
\right)
\\\nonumber
\times
\left(
\frac{\langle b|\textbf{r}|n\rangle^{*}\langle n |\textbf{r}|a\rangle^{*}}{E_{n}-E_{a}-\omega+\mathrm{i}\eta}+\frac{\langle b|\textbf{r}|n\rangle^{*} \langle n |\textbf{r}|a\rangle^{*}}{E_{n}-E_{b}+\omega+\mathrm{i}\eta}
\right)
n_{\beta}(\omega)n_{\beta}(|\omega_{ba}|-\omega)
,
\end{gather}
\begin{gather}
\label{GBp}
\Gamma_{ab}^{2\gamma,\mathrm{ram}}=\frac{8e^4}{9\pi}\frac{1}{2l_{a}+1}
\lim_{\eta\rightarrow 0}\mathrm{Re}
\sum\limits_{m_am_b}\int\limits_{|\omega_{ba}|}^{\infty}
d\omega
\omega^3(\omega-|\omega_{ba}|)^3
\sum\limits_{n}
\left(
\frac{\langle b|\textbf{r}|n\rangle \langle n |\textbf{r}|a\rangle}
{E_{n}-E_{a}-\omega+\mathrm{i}\eta}+
\frac{\langle b|\textbf{r}|n\rangle \langle n |\textbf{r}|a\rangle}
{E_{n}-E_{b}+\omega+\mathrm{i}\eta}
\right)
\\\nonumber
\times
\left(
\frac{\langle b|\textbf{r}|n\rangle^{*}\langle n |\textbf{r}|a\rangle^{*}}{E_{n}-E_{a}-\omega+\mathrm{i}\eta}+\frac{\langle b|\textbf{r}|n\rangle^{*} \langle n |\textbf{r}|a\rangle^{*}}{E_{n}-E_{b}+\omega+\mathrm{i}\eta}
\right)
(n_{\beta}(\omega)+n_{\beta}(\omega)n_{\beta}(\omega-|\omega_{ba}|))
,
\end{gather}
\begin{gather}
\label{GCp}
\Gamma_{ab}^{2\gamma,\mathrm{int}}=\frac{8e^4}{9\pi}\frac{1}{2l_{a}+1}
\lim_{\eta\rightarrow 0}\mathrm{Re}
\sum\limits_{m_am_b}\int\limits_{0}^{\infty}
d\omega
\omega^3(|\omega_{ba}|+\omega)^3
\sum\limits_{n}
\left(
\frac{\langle b|\textbf{r}|n\rangle \langle n |\textbf{r}|a\rangle}
{E_{n}-E_{a}+\omega+\mathrm{i}\eta}+\frac{\langle b|\textbf{r}|n\rangle \langle n |\textbf{r}|a\rangle}
{E_{n}-E_{b}-\omega+\mathrm{i}\eta}
\right)
\\\nonumber
\times
\left(
\frac{\langle b|\textbf{r}|n\rangle^{*}\langle n |\textbf{r}|a\rangle^{*}}
{E_{n}-E_{a}+\omega+\mathrm{i}\eta}+\frac{\langle b|\textbf{r}|n\rangle^{*} \langle n |\textbf{r}|a\rangle^{*}}
{E_{n}-E_{b}-\omega+\mathrm{i}\eta}
\right)
n_{\beta}(\omega)n_{\beta}(|\omega_{ba}|+\omega)
.
\end{gather}

From the expressions above follows that their structure is similar to the ordinary "zero-temperature" two-photon decay widths, see~\cite{ourtwoloop}. In absence of the resonant intermediate states in the sum over $ n $, Eqs.~(\ref{GAm}) and Eqs.~(\ref{GAp}), the imaginary infinitesimal part $ \mathrm{i}\eta $ in each energy denominator can be omitted. The situation is slightly different for the Raman-like contributions given by Eqs.~(\ref{GBm}), (\ref{GBp}). Since the integration interval over the frequency $ \omega $ in these equations is presented by the real half-axis, there are an infinite number of resonances for any initial and final states $ a $ and $ b $. Therefore, the contributions (\ref{GBm}) and (\ref{GBp}) can not be interpreted as the pure Raman scattering rate expressed by Eq.~(\ref{raman}). The same holds for the remaining interference contributions Eqs.~(\ref{GCm}) and (\ref{GCp}). Their algebraic structure allows one to validate that this is the interference contribution between Raman and emission (absorption) branches. 

Numerical calculation with the summation over entire spectrum in Eqs. (\ref{twophoton2})-(\ref{raman}) for the transition rates and thermal two-photon decay widths in Eqs.~(\ref{GAm})-(\ref{GCp}) were performed with the use of B-spline method~\cite{dkb}. The results for $ 2s $ and $ 3s $  states in H and He$ ^{+} $ atoms are presented in Tables~\ref{tab1}-\ref{tab4b}. 

\begin{table}
\caption{Different contributions to the partial two-photon decay width $ \Gamma^{2\gamma,\mathrm{BBR}}_{2s1s}  $ and the total two-photon decay width $ \Gamma_{2s}  $ (in s$ ^{-1} $) at different temperatures $ T $ (in Kelvin) in H. Values $ \Gamma_{2s1s}^{2\gamma,\mathrm{trans}}  $  in the first line coincides with the induced transition rates $W^{2\gamma,\mathrm{ind}}_{2s1s}  $, see Table~\ref{tab1}, since the cascades are absent for the partial widths. The partial contribution $ \Gamma_{2s1s}^{2\gamma,\mathrm{int}}  $ is negligibly small at all given temperatures.}
\begin{tabular}{ l c c c c c c }
\hline
\hline
$T$ & 77 & 300 & 1000 & 3000  & 5000  & 10$ ^{4} $ \\
\hline
$ \Gamma_{2s1s}^{2\gamma,\mathrm{trans}} $ & $1.358\times 10^{-4}$ & $2.028\times 10^{-3}$ &$ 2.151\times 10^{-2}$ & $ 1.731\times 10^{-1}$ & $4.389\times 10^{-1} $ & $1.467 $\\
$ \Gamma_{2s1s}^{2\gamma,\mathrm{ram}}   $ & $1.373\times 10^{-4}$ & $2.120\times 10^{-3}$ & $2.500\times 10^{-2}$ & $ 2.917\times 10^{-1}$ & $9.141\times 10^{-1}$ &$2.455$  \\
$ \sum_{b}\Gamma_{2sb}^{2\gamma,\mathrm{trans}}   $ &$ 1.358\times 10^{-4}$ & $2.028\times 10^{-3} $ & $2.151\times 10^{-2} $& $ 1.732\times 10^{-1}$ & $4.546\times 10^{-1}$   & $1.481$ \\
$ \sum_{b}\Gamma_{2sb}^{2\gamma,\mathrm{ram}}     $ & $ 1.373\times 10^{-4}$ & $2.120\times 10^{-3}$ & $2.500\times 10^{-2}$ & $2.916\times 10^{-1}$  & $ 9.039\times 10^{-1} $ & $ 2.395 $ \\
$ \sum_{b}\Gamma_{2sb}^{2\gamma,\mathrm{int}}     $ & $ 1.677\times 10^{-17} $ & $ 3.466\times 10^{-13} $ & $ 1.655\times 10^{-9} $ & $ 2.452\times 10^{-5} $  & $ 1.511\times 10^{-3} $ & $ 5.972\times 10^{-2} $ \\
$ \sum_{b}\Gamma_{2sb}^{2\gamma,\mathrm{trans+ram+int}}   $ & $ 2.373\times 10^{-4} $ & $ 4.148\times 10^{-3} $ & $ 4.651\times 10^{-2} $ & $   4.648\times 10^{-1} $ & $ 13.600\times 10^{-1} $ & $ 3.936 $ \\
\hline
\hline
\end{tabular}
\label{tab3}
\end{table}

\begin{table}
\caption{Different contributions to the partial two-photon decay width $ \Gamma^{2\gamma,\mathrm{BBR}}_{3s1s}  $ and the total two-photon decay width $ \Gamma_{3s}  $ (in s$ ^{-1} $) at different temperatures $ T $ (in Kelvin) in H. The partial contribution $ \Gamma_{3s1s}^{2\gamma,\mathrm{int}}  $ is negligibly small at all given temperatures. The zero-temperature two-photon widths is $\Gamma^{2\gamma}_{3s1s}= 2.082854$ s$ ^{-1} $.}
\begin{tabular}{ l c c c c c c }
\hline
\hline
$T$ & 77 & 300 & 1000 & 3000  & 5000  & 10$ ^{4} $ \\
\hline
$ \Gamma_{3s1s}^{2\gamma,\mathrm{trans}}                  $ & $ 2.161\times 10^{-4} $ & $ 3.179\times 10^{-3} $ & $ 3.268\times 10^{-2} $ & $ 2.632\times 10^{-1} $ & $ 6.707\times 10^{-1} $ & $ 1.924 $\\
$ \Gamma_{3s1s}^{2\gamma,\mathrm{ram}}                    $ & $ 2.214\times 10^{-4} $ & $ 3.498\times 10^{-3} $ & $ 4.751\times 10^{-2} $ &$ 3.980\times 10^{-1} $  & $ 6.675\times 10^{-1} $  & $ 6.822\times 10^{-1} $\\
$ \sum_{b}\Gamma_{3sb}^{2\gamma,\mathrm{trans}}           $ & $ 2.382\times 10^{-4} $ & $ 3.500\times 10^{-3} $  & $ 3.585\times 10^{-2} $ & $ 2.863\times 10^{-1} $ & $ 7.253\times 10^{-1} $  & $ 2.116 $ \\
$ \sum_{b}\Gamma_{3sb}^{2\gamma,\mathrm{ram}}             $ & $ 2.442\times 10^{-4} $ & $ 3.863\times 10^{-3} $ & $ 5.275\times 10^{-2} $ & $ 4.390\times 10^{-1}  $ & $ 7.386\times 10^{-1} $ & $ 9.513\times 10^{-1} $\\
$ \sum_{b}\Gamma_{3sb}^{2\gamma,\mathrm{int}}     $ & $ 1.285\times 10^{-15} $ & $ 2.731\times 10^{-11}$ & $5.862\times 10^{-7} $ & $ 8.614\times 10^{-4} $  & $ 7.269\times 10^{-3} $ & $8.337\times 10^{-2} $ \\
$ \sum_{b}\Gamma_{3sb}^{2\gamma,\mathrm{trans+ram+int}}   $ & $ 4.824\times 10^{-4} $ & $ 7.363\times 10^{-3} $ & $ 8.860\times 10^{-2} $ & $  7.261\times 10^{-1}$ & $ 1.471 $ & $ 3.151 $\\
\hline
\hline
\end{tabular}
\label{tab3b}
\end{table}

\begin{table}
\caption{Different contributions to the partial two-photon decay width $ \Gamma^{2\gamma,\mathrm{BBR}}_{2s1s}  $ and the total thermal two-photon decay width $ \Gamma_{2s}  $ (in s$ ^{-1} $) at different temperatures $ T $ (in Kelvin) in He$ ^{+} $. The partial contribution $ \Gamma_{2s1s}^{2\gamma,\mathrm{int}}  $ is negligibly small at all given temperatures.}
\begin{tabular}{ l c c c c c c  }
\hline
\hline
$T$ & 77 & 300 & 1000 & 3000  & 5000  & 10$ ^{4} $ \\
\hline
$ \Gamma_{2s1s}^{2\gamma,\mathrm{trans}}     $ & $ 5.452\times 10^{-4} $ & $ 8.243\times 10^{-3} $ & $ 9.044\times 10^{-2} $ & $ 7.868\times 10^{-1} $ & $ 2.118 $ & $ 7.893 $ \\
$ \Gamma_{2s1s}^{2\gamma,\mathrm{ram}}     $ & $ 5.468\times 10^{-4} $ & $ 8.335\times 10^{-3} $ & $ 9.385\times 10^{-2} $ & $ 8.801\times 10^{-1} $ & $ 2.562 $ &  $ 12.072 $ \\
$ \sum_{b}\Gamma_{2sb}^{2\gamma,\mathrm{trans}}     $ & $ 5.452\times 10^{-4} $ & $ 8.243\times 10^{-3} $  & $ 9.044\times 10^{-2} $ & $ 7.868\times 10^{-1} $ & $ 2.118 $ & $ 7.894 $  \\
$ \sum_{b}\Gamma_{2sb}^{2\gamma,\mathrm{ram}} $ & $ 5.468\times 10^{-4} $ & $ 8.335\times 10^{-3}  $ & $ 9.385\times 10^{-2}  $ & $ 8.801\times 10^{-1} $ & $ 2.562 $ & $ 12.08 $ \\
$ \sum_{b}\Gamma_{2sb}^{2\gamma,\mathrm{int}}     $ & $6.546\times 10^{-20} $ & $ 1.351\times 10^{-15}$ & $6.183\times 10^{-12} $ & $1.384\times 10^{-8} $  & $5.241\times 10^{-7} $ & $ 2.558\times 10^{-4}$ \\
$ \sum_{b}\Gamma_{2s}^{2\gamma,\mathrm{trans+ram+int}} $ & $ 21.840\times 10^{-4} $ & $ 1.658\times 10^{-2} $ & $ 1.843\times 10^{-1} $ & $ 1.667 $ & $ 2.774  $ & $ 19.974 $\\
\hline
\hline
\end{tabular}
\label{tab4}
\end{table}

\begin{table}
\caption{Different contributions to the partial two-photon decay width $ \Gamma^{2\gamma,\mathrm{BBR}}_{3s1s}  $ and the total thermal two-photon decay width $ \Gamma_{3s}  $ (in s$ ^{-1} $) at different temperatures $ T $ (in Kelvin) in He$ ^{+} $. The partial contribution $ \Gamma_{3s1s}^{2\gamma,\mathrm{int}}  $ is negligibly small at all given temperatures. The zero-temperature two-photon widths is $\Gamma^{2\gamma}_{3s1s}= 1.333\times 10^2$ s$ ^{-1} $.}
\begin{tabular}{ l c c c c c c  }
\hline
\hline
$T$ & 77 & 300 & 1000 & 3000  & 5000  & 10$ ^{4} $ \\
\hline
$ \Gamma_{3s1s}^{2\gamma,\mathrm{trans}}     $ & $ 8.722\times 10^{-4} $ & $ 1.312\times 10^{-2} $ & $ 1.422\times 10^{-1} $ & $ 1.206 $ & $ 3.197 $ & $ 11.90 $ \\
$ \Gamma_{3s1s}^{2\gamma,\mathrm{ram}}        $ & $ 8.772\times 10^{-4} $ & $ 1.344\times 10^{-2} $ & $ 1.540\times 10^{-1} $ & $ 1.571 $ & $ 5.088 $ & $ 19.610 $ \\ 
$ \sum_{b}\Gamma_{3sb}^{2\gamma,\mathrm{trans}}     $ & $ 9.615\times 10^{-4} $ & $ 1.447\times 10^{-2} $  & $ 1.566\times 10^{-1}  $ & $ 1.325$ & $ 3.504 $ & $ 12.97 $  \\
$ \sum_{b}\Gamma_{3sb}^{2\gamma,\mathrm{ram}} $ & $ 9.674\times 10^{-4} $ & $ 1.482\times 10^{-2} $ & $ 1.700\times 10^{-1}$ & $ 1.741 $ & $5.652 $ & $ 21.670  $ \\
$ \sum_{b}\Gamma_{3sb}^{2\gamma,\mathrm{int}}     $ & $ 4.997\times 10^{-18}$ & $ 1.031\times 10^{-13}$ & $ 4.823\times 10^{-10} $ & $ 2.267\times 10^{-6} $  & $ 2.727\times 10^{-4} $ & $ 2.313\times 10^{-2} $ \\
$ \sum_{b}\Gamma_{3s}^{2\gamma,\mathrm{trans+ram+int}} $ & $ 1.929\times 10^{-3} $ & $ 2.929\times 10^{-2} $ & $ 3.266\times 10^{-1} $ & $3.066  $ & $ 9.156 $ & $ 34.663 $\\
\hline
\hline
\end{tabular}
\label{tab4b}
\end{table}

\section{Discussion and conclusions}
\label{theend}

In this paper, we have considered thermal two-loop self-energy corrections. Their imaginary part represents the two-photon thermal correction to the natural (spontaneous) width of the atomic energy level. In general, the total contribution can be expressed by Eqs. (\ref{totalg})-(\ref{GCp}). Then, according to Eqs. (\ref{totalwidths})-(\ref{finitet}) the contribution $\Gamma_{a}^{\mathrm{BBR}}$ should be compared with others "ordinary" radiative QED corrections and with the BBR-induced rates known from the quantum mechanical approach.

The relativistic and "zero-temperature" QED corrections, see \cite{benidikt}, of the leading order to the $ 2s $ level width in the nonrecoil limit are given by 
\begin{eqnarray}
\label{corr}
\widetilde{W}^{2\gamma,\mathrm{spon}}_{ab}=W^{2\gamma,\mathrm{spon}}_{ab}
\left[ 
1+\epsilon^{\mathrm{rel}} + \epsilon^{\mathrm{QED}}
\right]
,
\end{eqnarray}
where 
\begin{eqnarray}
\label{rel}
\epsilon^{\mathrm{rel}}=c_{2}(\alpha Z)^2, 
\end{eqnarray}
\begin{eqnarray}
\label{qed}
\epsilon^{\mathrm{QED}}=c_{3}\frac{\alpha}{\pi}(\alpha Z)^2\mathrm{ln}[(\alpha Z)^{-2}]
.
\end{eqnarray}
The coefficients $ c_{2} $ and $ c_{3} $ in the equation above were evaluated for the two-photon decay rate of higher excited $ ns $ and $ nd $ states in~\cite{benidikt}. Using numerical values of the parameters $ c_2 $ and $ c_3  $ one can find that the relativistic and radiative QED corrections to the $ 2s\rightarrow 1s+\gamma(\mathrm{E1}) $ transition in H atom are equal to $ -2.908\times 10^{-4} $ s$^{-1 } $  and $ -2.024\times 10^{-5} $ s$^{-1 } $, respectively. The depopulation rate with the account for the finite lifetimes and thermal two-photon level widths of the $ 2s $ state at room temperature are $\Gamma_{2s}^{1\gamma,\mathrm{BBR-QED}}= 4.159 \times 10^{-3} $ s$^{-1 } $ and $\sum_{b}\Gamma_{2sb}^{2\gamma,\mathrm{trans+ram+int}}=4.148\times 10^{-3}  $ s$^{-1 } $, see Tables~\ref{tab_comparison} and \ref{tab3}. Thus, the two-loop thermal corrections to the $ 2s $ level broadening in H atom dominate over the relativistic and radiative corrections. However, all these corrections are still less than the experimental uncertainty in measuring the lifetime of the $ 2s $ state in hydrogen, see Table~\ref{tab5}.

The situation is different for the $ 2s $ state in the He$^+$ atom, where more accurate experiments were carried out, and the measured decay rate is $ 525\pm 5 $ s$^{-1 } $~\cite{hinds}. According to Eq.~(\ref{corr}) the relativistic and radiative QED corrections to the $ 2s-1s $ transition in He$ ^{+} $ atom are $ -7.445\times 10^{-2} $ s$^{-1 } $  and $ -4.451\times 10^{-3} $, respectively. Assuming that the experiment \cite{hinds} is carried out at the room temperature, one can find the corresponding BBR-induced depopulation rate $\Gamma_{2s}^{1\gamma,\mathrm{BBR-QED}}=1.720  \times 10^{-2} $ s$^{-1 }$, and total thermal two-photon width  $\sum_{b}\Gamma_{2sb}^{2\gamma,\mathrm{trans+ram+int}}=1.658\times 10^{-2}$ s$^{-1 }$, see Table \ref{tab4}. Again, the two-loop thermal corrections to the $ 2s $ level broadening at $ T=300 $ K in He$ ^{+} $ ion reach the same order as the one-photon depopulation width at higher temperatures. Although, the contribution of thermal corrections is two orders less than the experimental uncertainty, the further improvement of accuracy could verify directly this effect. The current status of experiments on the measurement of $2s$ state lifetime in other H-like ions is presented in Table~\ref{tab5}. 

Except the He$^{+}$ results, the precision on the level of one percent is also achieved for Ar$^{17+}$ and Ni$^{27+}$ hydrogen-like ions. However, since the Planck distribution function is mainly in the low-frequency region at low temperatures one can neglect $ \omega $ in the energy denominators of Eqs. (\ref{GAm})-(\ref{GCp}). This leads to the parametric estimation $ (k_{B}T)^4/(m^3Z^2) $ r.u. for the $ \Gamma^{2\gamma,\mathrm{BBR}} $ contribution in the low temperature regime. Taking in mind that $ k_{B}T\sim m(\alpha Z)^2 $ in r.u., the well-known $\alpha Z$-parametrization can be found in the form $ m\alpha^2(\alpha Z)^6 $ for two-photon contribution $\Gamma^{2\gamma,\mathrm{BBR}} $. From this it follows that $ \Gamma^{2\gamma,\mathrm{BBR}} $ behaves as $ Z^6 $. Using this estimation, numerical calculations show that the thermal effect described in this paper is much less than the uncertainty of experiments listed in Table~\ref{tab5}.

It is also interesting to note the peculiarity of the total Raman-like contribution $ \sum_{b}\Gamma_{ab}^{2\gamma,\mathrm{ram}}  $ to the two-photon level broadening. As seen from the Table \ref{tab3} the total value $ \sum_{b}\Gamma_{ab}^{2\gamma,\mathrm{ram}}  $ at some temperatures become less than the partial contribution $ \Gamma_{ab}^{2\gamma,\mathrm{ram}}  $. This occurs via the contributions with $ b>a $ in the sum over $ b $ in Eq. (\ref{GBp}). They change the sign with the increasing of temperature and, as a result, this leads to the decreasing of the total value. The same situation is well-known for the thermal Stark-shift, see for example \cite{solovyev2015,farley}. Such behavior of partial contributions, namely that they can be negative, was found previously in the work \cite{ourtwoloop}. In addition, the numerical calculations showed that all two-photon contributions proportional to $ n_{\beta}(\omega)n_{\beta}(\omega')$ are negligible at the reasonable temperatures. In particular, it was found that the interference part given by Eqs. (\ref{GCm}) and (\ref{GCp}), as well as the absorption-like part given by Eq. (\ref{GAp}), does not make a tangible contribution to the thermal two-photon width.

The corrections found in this work can also play an important role in description the cosmological recombination processes. The standard calculation of the ionization fraction of the primordial hydrogen plasma takes into account only the induced two-photon decay and Raman scattering. In \cite{chlubaind,hirataind,kholupenkoind} it was shown that the induced two-photon transition $ 2s\rightarrow 1s+2\gamma $ corrects the ionization fraction on the level of a few percent. The two-loop thermal corrections for the $ 2s $ level width listed in Table~\ref{tab3} are several times larger then the corresponding induced rates at recombination temperatures ($ 1000<T<5000 $). Thus, one can expect the contribution to the ionization fraction at the same level. 

\begin{table}[hbtp]
\caption{Comparison of the experimental and theoretical lifetimes $ \tau_{2s}=\Gamma^{-1}_{2s} $ (in s) of the $2s$ level in H-like ions. Theoretical values of $ \tau_{2s} $ were calculated in a fully relativistic approach with the account for finite nuclear size effects. The M1 decay rate $ 2s\rightarrow 1s +\gamma(\mathrm{M1}) $ is also taken into account.}
\begin{tabular}{c l c }
\hline
\hline
$ Z $ & Experiment (s) & Theory (s) \\
\hline
 1 & $0.67\pm 0.29$ $^{a}$& $1.216\times 10^{-1}$ \\
   & $0.12^{+0.03}_{-0.04} $ $^{b}$& \\ 
 2 & $(1.922\pm 0.082)\times 10^{-3}$ $^{c}$& $1.898\times 10^{-3}$ \\ 
   & $(2.04^{+0.81}_{-0.34})\times 10^{-3}$ $ ^{d} $& \\
   & $(1.905\pm 0.018)\times 10^{-3}$ $^{e}$& \\
 8 & $(4.53\pm 0.43)\times 10^{-7}$ $^{f}$& $ 4.636\times 10^{-7}$\\
 9 & $(2.37\pm 0.19)\times 10^{-7}$ $^{f}$& $2.286\times 10^{-7} $ \\
 16 & $(7.3\pm 0.7)\times 10^{-9}$ $^{g}$& $7.153\times 10^{-9}$ \\
 18 & $(3.54\pm 0.25)$ $\times 10^{-9}$ $^{g}$& $3.494\times 10^{-9}$ \\
    & $(3.487\pm 0.036)$ $\times 10^{-9}$ $^{h}$&   \\
 28 & $(2.171\pm 0.018)$ $\times 10^{-10}$ $^{i}$& $2.156\times 10^{-10} $ \\
 36 & $(3.68\pm 0.14)$ $\times 10^{-11}$ $^{j}$& $3.701\times 10^{-11}$\\
\hline
\hline
\end{tabular}
\\
\begin{flushleft}
$^a $ Reference~\cite{kruger}, $^b $ Reference~\cite{2slife} (this value is extracted from the analysis of CMB), $^c $ Reference~\cite{prior}, $^e $ Reference~\cite{hinds}, $^f $ Reference~\cite{cocke}, $^g $ Reference~\cite{marrus}, $^h $ Reference~\cite{gould}, $^i $ Reference~\cite{dunford}, $^j $ Reference~\cite{cheng}

\end{flushleft}
\label{tab5}
\end{table}

In conclusion, we can expect that the radiative temperature-dependent corrections to the level widths found in this paper will play a role in both astrophysical and laboratory investigations. 
At least, a further increase of the experimental accuracy faces the need to take these corrections into account. 

\section*{Acknowledgements}
This work was supported by Russian Science Foundation (Grant No. 17-12-01035).

\end{document}